\documentclass[reprint, amsmath,amssymb, aps, pra]{revtex4-1}

\usepackage{graphicx}
\usepackage{dcolumn}
\usepackage{bm}

\begin{document}

\title{Isotopic variation of parity violation in atomic ytterbium: method of measurements and analysis of systematic effects}
 
\author{D. Antypas}
\email{dantypas@uni-mainz.de}
\affiliation{Helmholtz-Institut Mainz, Mainz 55128, Germany}

\author{A.M. Fabricant}
\affiliation{Johannes Gutenberg-Universit{\"a}t Mainz, Mainz 55128, Germany}

\author{J.E. Stalnaker}
\affiliation{Department of Physics and Astronomy, Oberlin College, Oberlin, Ohio 44074, USA}

\author{K. Tsigutkin}
\affiliation{ASML, Veldhoven, The Netherlands}

\author{V.V. Flambaum}
\affiliation{
School of Physics, University of New South Wales, Sydney 2052, Australia}
\affiliation{Johannes Gutenberg-Universit{\"a}t Mainz, Mainz 55128, Germany}

\author{D. Budker}
\affiliation{Johannes Gutenberg-Universit{\"a}t Mainz, Mainz 55128, Germany}
\affiliation{Helmholtz-Institut Mainz, Mainz 55128, Germany}
\affiliation{Department of Physics, University of California at Berkeley, California 94720-300, USA}

\date{\today}

\begin{abstract}
We present a detailed description of experimental studies of the parity violation effect in an isotopic chain of atomic ytterbium (Yb), whose results were reported in a recent Letter [Antypas \textit{et al.}, Nat. Phys. 15, 120 (2019) \cite{Antypas2019IsotopicYtterbium}]. We discuss the principle of these measurements, made on the Yb 6s$^2$ $^1$S$_0 \rightarrow $5d6s $^3$D$_1$ optical transition at 408 nm, describe the experimental apparatus, and give a detailed account of our studies of systematic effects in the experiment. Our results offer the first direct observation of the isotopic variation in the atomic parity violation effect, a variation which is in agreement with the prediction of the Standard Model. These measurements are used to constrain  electron-proton and electron-neutron interactions, mediated by a light $Z'$ boson.   

\begin{description}

\item[PACS numbers]
 \verb+11.30.Er, 32.90.+a+

\end{description}
\end{abstract}

\pacs{Valid PACS appear here}

\maketitle

\bibliographystyle{apsrev4-1}

\section{\label{sec:level1-Introduction}INTRODUCTION}
The investigation of weak-force-induced effects in atomic systems has been the focus of experiments in the last four decades (see, for example, reviews \cite{Ginges2004ViolationsParticles, Roberts2015ParitySystems, Safronova2018SearchMolecules}). The first experiments were motivated by the work of Bouchiat and Bouchiat \cite{Bouchiat1974WeakPhysics} which showed that weak-interaction-induced observables in atoms are enhanced and therefore are detectable in systems with large atomic number. This finding followed the earlier recognition by Zel'dovich \cite{ZelDovich1959PARITYEFFECTS} that the electron-nucleus weak interaction induces optical rotation in atomic media. Atomic physics techniques have been employed to study the parity violation (PV) at low energy. Combined with atomic structure calculations, these efforts have determined the nuclear weak charge, a quantity predicted in the Standard Model (SM), thereby testing the SM. Such tabletop experiments are complementary to studying the electroweak sector of the SM at high energies. 

The first observations of atomic PV were made in bismouth (Bi) \cite{Barkov1978ObservationTransitions}, thallium (Th) \cite{Conti1979PreliminaryThallium} and cesium (Cs) \cite{Bouchiat1982ObservationCesium}.
Accurate determinations of the PV effects were made in Bi \cite{MacPherson1991PreciseBismuth}, lead (Pb) \cite{Meekhof1993High-precisionLead, Phipp1996ALead}, Th \cite{Vetter1995PreciseThallium, Edwards1995PreciseThallium} and Cs \cite{Wood1997MeasurementCesium, Guena2005MeasurementDetection}. The highest measurement accuracy was achieved in Cs  \cite{Wood1997MeasurementCesium}. Combined with precise atomic-structure calculations \cite{Dzuba2012RevisitingCesium}, the Cs experiment resulted in a determination of the nuclear weak charge at the level of 0.5\%.  This result is the most-precise-to-date low-energy test of the SM.

Atomic PV experiments can additionally be platforms to study nuclear physics as well as physics beyond the SM. Measurements of nuclear-spin-dependent contributions to the PV effect  probe the nuclear anapole moment \cite{Flambaum1980V.V.1980, Flambaum1980V.V.52-1980, V.V.Flambaum1984PossibilityExperiments}, which has only been observed to date in the Cs experiment \cite{Wood1997MeasurementCesium}. Determining an anapole provides information about the so-far poorly understood weak meson-nucleon couplings  that characterize the hadronic weak interactions, as formulated in the model of Desplanques, Donoghue, and Hollstein \cite{Desplanques1980UnifiedForce}.  Measurements of PV across a chain of isotopes of the same element, first proposed in \cite{Dzuba1986EnhancementAtoms}, have the potential probe to physics beyond the SM \cite{Brown2009CalculationsViolation,Viatkina2019DependencePhysics}, such as to search for extra light bosons that mediate parity-violating interactions between the electron and nucleons \cite{Dzuba2017ProbingMolecules}. The isotopic comparison method can be also employed to probe the variation of the neutron distribution in the nucleus, and to test nuclear models  \cite{Fortson1990Nuclear-structureNonconservation, Viatkina2019DependencePhysics}.
 
A number of PV experiments are currently underway, that make use of neutral atoms, as well as atomic ions and molecules. Of these, an experiment in Fr \cite{Zhang2016EfficientAtoms} is aiming to determine the nuclear weak charge, as well as to measure the anapole moment of Fr nuclei. Another project with Fr, currently at a preliminary stage \cite{Aoki2017Parity-nonconservingModel}, also aims to measure the weak charge and anapole. An experiment using a single trapped Ra$^+$ ion \cite{NunezPortela2014Ra+Clock}, aims to determine the nuclear weak charge in several different isotopes. An ongoing experiment in Cs  \cite{Choi2016MeasurementControl} is primarily focused on a cross-check measurement of the Cs anapole moment. Improved measurements of PV are underway in Dy \cite{Leefer2014TowardsDysprosium}, in which a previous experiment yielded an effect consistent with zero \cite{Nuyen1997SearchDysprosium}. Finally, an effort with BaF \cite{Altuntas2018DemonstrationViolation, Altuntas2018MeasuringErrors} has recently demonstrated adequate sensitivity to make an accurate determination of the anapole moment of the Ba nucleus.

Accurate extraction of the nuclear weak charge from PV measurements requires atomic calculations of adequate precision.  Such a precision can be reached in simple atomic systems such as Cs, Fr or Ra$^+$ (the Cs theory, for example, is at the 0.5\% level of uncertainty \cite{Dzuba2012RevisitingCesium}), thus making it possible for a single-isotope measurement to be a probe of the SM. In Yb, which has two valence electrons, existing atomic calculations have a relatively large uncertainty at the 10\% level \cite{Porsev1995ParityYtterbium,Dzuba2011CalculationYtterbium}. Significant advancement in the Yb theory is required to enable a competitive determination of the Yb weak charge. With regard to searching for physics beyond the SM via atomic PV, the merit of using Yb lies in the availability of a number of stable isotopes, that makes it possible to employ the isotopic comparison method \cite{Dzuba1986EnhancementAtoms}. The same method could also be used to probe the neutron distributions of the Yb nuclei.

We recently reported on measurements of PV in the 6s$^2$ $^1$S$_0$ $\rightarrow$ 5d6s $^3$D$_1$ optical transition at 408 nm in a chain of four nuclear-spin-zero Yb isotopes \cite{Antypas2019IsotopicYtterbium}.  That work provided an observation of the isotopic variation of the PV effect, and was part of a program that focuses on nuclear spin-dependent PV, neutron skins, as well as on searching for light bosons beyond SM.  These results built upon an earlier observation of the Yb PV effect \cite{Tsigutkin2009ObservationYtterbium, Tsigutkin2010ParitySystematics}. The previous measurement confirmed the large size of the effect, which was first estimated in \cite{DeMille1995ParityYtterbium}, with more elaborate calculations following up \cite{Porsev1995ParityYtterbium, Das1997ComputationYtterbium,Dzuba2011CalculationYtterbium}.  Here we present in detail the method utilized for these isotopic-chain measurements, discuss the experimental apparatus, and provide an analysis of systematic effects.

\section{\label{sec:level1-Experimental Method}Experimental Method}

To study PV in Yb, we make use of the 6s$^2$ $^1$S$_0$ $\rightarrow$ 5d6s $^3$D$_1$ optical transition at 408 nm (fig. 1). The experimental principle was described in \cite{Tsigutkin2010ParitySystematics}. A small electric-dipole (\textit{E}1) transition amplitude arises between the $^1$S$_0$ and $^3$D$_1$ states, mainly due to weak-interaction-induced mixing between the $^3$D$_1$ and $^1$P$_1$ states. The application of a quasi-static electric field creates additional (Stark) mixing between the same states \cite{Bouchiat1975ParityII}, and introduces a Stark-induced \textit{E}1 amplitude for the 408 nm transition. A static magnetic field is also applied to the atoms to split the Zeeman sublevels of the excited $^3$D$_1$ state. With appropriate choice of geometry for the applied static and optical fields, the Stark and PV amplitudes interfere \cite{Bouchiat1986OpticalInteractions}. The sign of this interference in the 408 nm excitation rate can be changed by making field reversals, allowing extraction of the P-odd part of this rate from the larger P-even background. For the geometry of fields in the present experiment (fig. \ref{fig:apparatus}), the Stark-PV interference is proportional to the following pseudo-scalar rotational invariant \cite{Bouchiat1986OpticalInteractions, Drell1984ParityThallium}:
\begin{equation}
\label{eq:RotInvariant}
(\vec{\mathcal{E}}\cdot\vec{B})\cdot([\vec{E}\times\vec{\mathcal{E}}]\cdot\vec{B}),
\end{equation}

\noindent where $\vec{E}$, $\vec{\mathcal{E}}$ and $\vec{B}$ are, respectively, the quasi-static electric, optical and magnetic fields applied to the atoms. The Stark and PV amplitudes for the \textit{m}=0 $\rightarrow m'$ component of the $^1$S$_0$ $\rightarrow$$^3$D$_1$ transition are given by \cite{Tsigutkin2010ParitySystematics}:

\begin{equation}
\label{eq:Astark0}
A^{Stark}_{m'}=i\beta(-1)^{m'}(\vec{E}\times\vec{\mathcal{E}})_{-m'},
\end{equation}
\begin{equation}
\label{eq:APV0}
A^{PV}_{m'}=i\zeta(-1)^{m'}\vec{\mathcal{E}}_{-m'},
\end{equation}

\begin{figure}
\includegraphics[]{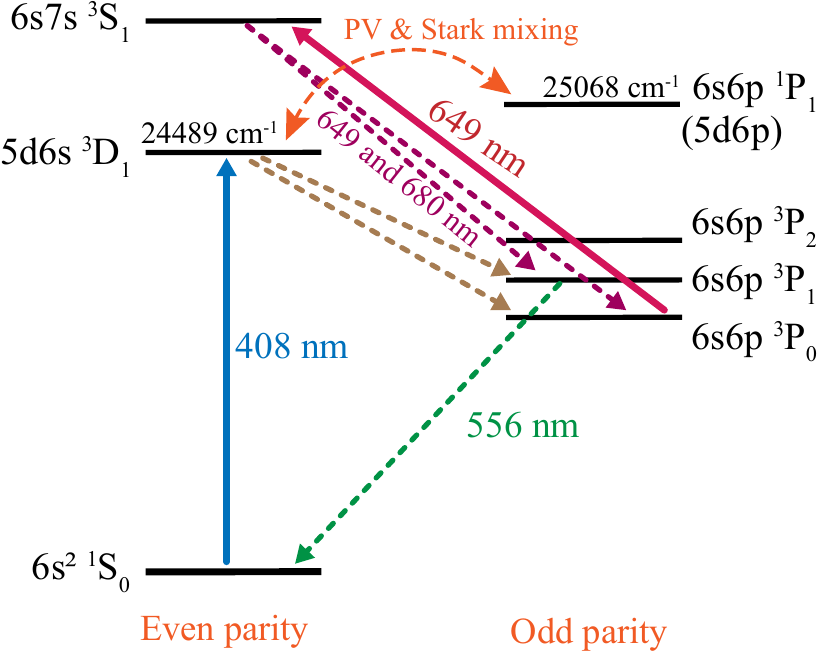}
\caption{(Color online)  Partial energy level diagram  of Yb with levels related to the PV experiment. Solid straight lines indicate excitations and dashed straight lines indicate decays. The PV effect arises primarily due to weak-interaction-induced mixing of the $^3$D$_1$ and $^1$P$_1$ levels. About 67\% of atoms excited to the  $^3$D$_1$ level decay to the metastable $^3$P$_0$ level. These atoms are detected by subsequent excitation to $^3$S$_1$ and collection of fluorescence from decays at 556, 649 and 680 nm.}
\label{fig:energylevels}
\end{figure}

\noindent where $\beta=2.24(12)\cdot10^{-8} ea_0$ /(V/cm) is the vector polarizability of the transition, determined in \cite{Bowers1999ExperimentalYtterbium,Stalnaker2006DynamicShapes}, and $\zeta$ is the \textit{E}1 transition moment arising primarily from the PV-mixing of the $^3$D$_1$ and $^1$P$_1$ states. The parameter $\zeta$ is proportional to the nuclear weak charge. The element $V_q$ is the \textit{q}-component of the vector $\vec{V}$ in the spherical basis. The results presented here come from measurements on the $m=0 \rightarrow m'=0$ transition component, whereas the previous experiment \cite{Tsigutkin2009ObservationYtterbium,Tsigutkin2010ParitySystematics} utilized all three magnetic sublevels of the 408 nm transition to determine the PV-effect. 

The effects of a magnetic-dipole (\textit{M}1) transition between the $^1$S$_0$ and $^3$D$_1$ states, whose amplitude is $\approx$930 times greater than that of the PV amplitude, are suppressed in this experiment. The primary method of suppression is the appropriate choice of the geometry of fields in the interaction region. This geometry is chosen such that the Stark and PV amplitudes are in phase and therefore allowed to interfere, but the \textit{M}1 and Stark amplitudes are nominally out of phase and do not interfere. As a result, the $M1$-related systematic contributions to the PV measurements are practically eliminated.  Additional suppression of $M1$-systematics occurs because the $^1$S$_0\rightarrow ^3$D$_1$ excitation is done with a standing-wave optical field. Analysis of the residual contribution of the $M1$ transition to the present measurements is carried out in Appendix \ref{sec:level1-AppendixA}.

In the absence of non-reversing fields and field misalignments, the magnetic field is along the z-axis, $\vec{B} =B_z\hat{z}$, and the electric field along the x-axis, $\vec{E} =(E_{dc} +E_0$cos$\omega t )\hat{x}$. This field consists of a component oscillating at frequency $\omega$ ($\omega/2\pi$=19.9 Hz) as well as a dc-term. The ac-component, of typical amplitude 1.2 kV/cm, is primarily responsible for the required Stark-induced mixing between $^3$D$_1$ and $^1$P$_1$ states. The change of the ac-field direction is the primary parity reversal in the experiment. The dc term \mbox{ $E_{dc}$ ($\approx$ 6V/cm)} is used to optimize detection conditions for the Stark-PV interference. The optical field is linearly polarized and propagates along x: $\vec{\mathcal{E}} =\mathcal{E}($sin$\theta \hat{y}$+cos$\theta \hat{z})$. Under these conditions, the excitation rate for the $m=0 \rightarrow m'$ transition component has the form:

\begin{equation}
\label{eq:R0}
R_{m'}\propto\vert A_{m'}^{Stark}+A_{m'}^{PV}\vert ^2\\
=R_{m'}^{[0]}+R_{m'}^{[1]}\cos\omega t +R_{m'}^{[2]}\cos2\omega t.
\end{equation}
\noindent This rate consists of a dc term of amplitude $R_{m'}^{[0]}$ and components oscillating at  frequencies $\omega$ and $2\omega$ with respective amplitudes $R_{m'}^{[1]}$ and $R_{m'}^{[2]}$. For the $0\rightarrow0$ transition these terms are as follows:
\begin{multline}
\label{eq:R0_zero_harm}
R_0^{[0]}=2\mathcal{E}^2\beta^2E_0^2\sin^2\theta+4\mathcal{E}^2\beta^2E_{dc}^2\sin^2\theta \\+8\mathcal{E}^2\beta E_{dc}\zeta\cos\theta\sin\theta,
\end{multline}
\begin{equation}
\label{eq:R1}
R_0^{[1]}=8\mathcal{E}^2\beta E_0\zeta\cos\theta \sin\theta+8\mathcal{E}^2\beta^2 E_0E_{dc}\sin^2\theta,
\end{equation}
\begin{equation}
\label{eq:R2}
R_0^{[2]}=2\mathcal{E}^2\beta^2 E_0^2\sin^2\theta.
\end{equation}
\noindent Only terms independent of or linear in the weak-interaction parameter $\zeta$ are retained in (\ref{eq:R0}), (\ref{eq:R1}) and (\ref{eq:R2}). Phase-sensitive detection at the frequencies $\omega$ and 2$\omega$  provides the amplitudes   $R_0^{[1]}$ and $R_0^{[2]}$. Their ratio is related to the ratio of the PV- and Stark-induced transition moments:
\begin{equation}
\label{eq:r0ideal}
r_0\equiv \frac{R_0^{[1]}}{R_0^{[2]}}=\frac{4E_{dc}}{E_0}+\frac{4\zeta}{\beta E_0}\cot\theta. 
\end{equation}
\noindent Observation of the change in $r_0$ under the second parity reversal, i.e. a $\pm\pi$/2 rotation of the light polarization plane, yields the ratio $\zeta$/$\beta$. In addition to the \textit{E}- and $\theta$- reversals (parity reversals), the magnetic field $\vec{B}$ as well as the polarity of $E_{dc}$ are also reversed, in order to study and minimize systematic contributions, not explicitly shown in (\ref{eq:r0ideal}). 

\begin{figure}
\includegraphics[]{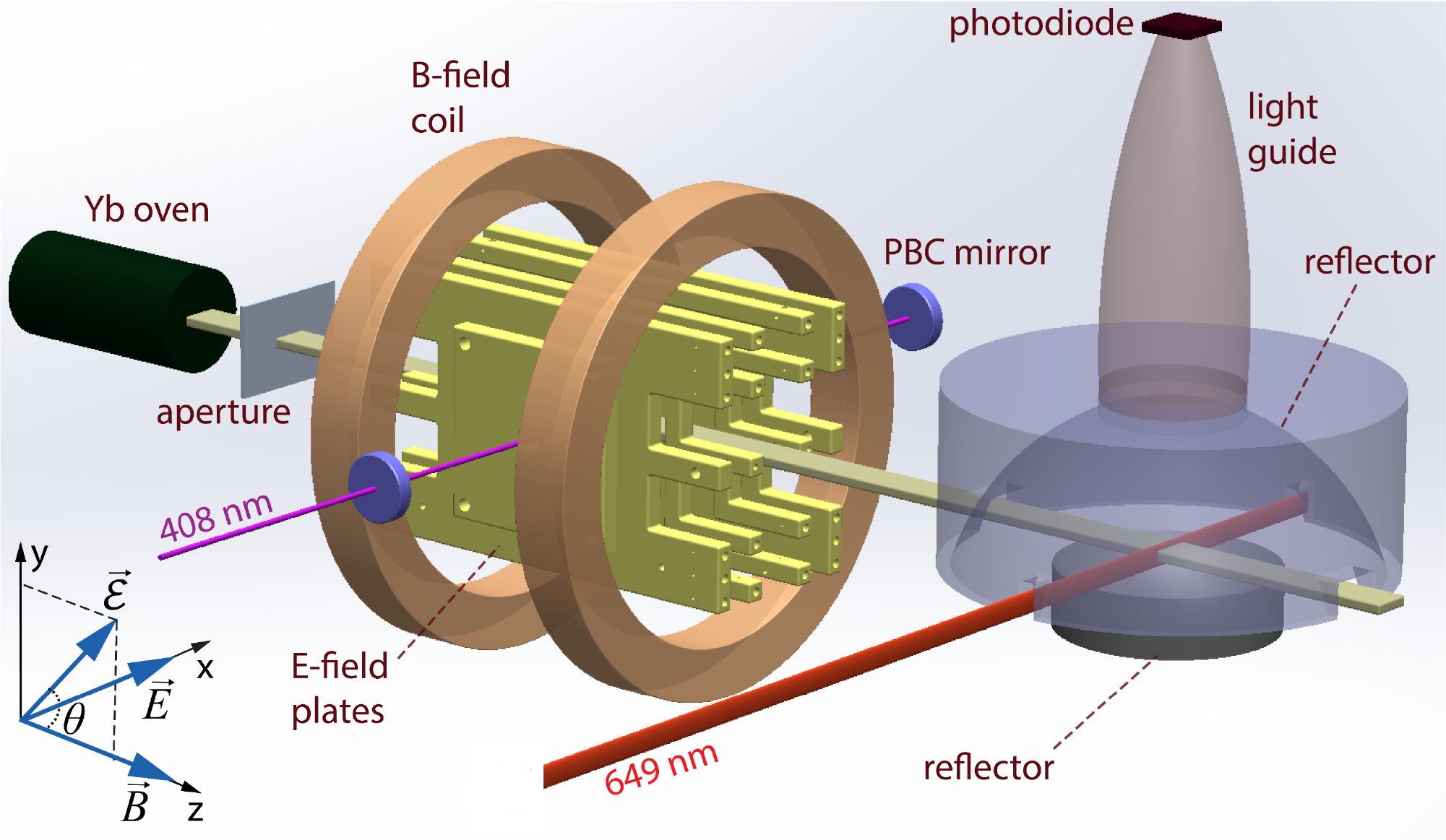}
\caption{(Color online) Schematic of the Yb atomic-beam apparatus. The Yb atoms effuse from the oven into the interaction region, where they are excited by 408 nm light in the presence of applied electric and magnetic fields.  The atoms that are excited are detected in the downstream detection region via excitation from the metastable state at 649 nm. Figure reproduced from \cite{Antypas2019IsotopicYtterbium}.}
\label{fig:apparatus}
\end{figure}

Misalignments of the applied fields, non-reversing field components, as well as imperfections in the optical polarization alter the ideal situation discussed above, and result in additional contributions to the transition rate (\ref{eq:R0}) and to the ratio (\ref{eq:r0ideal}). The applied electric field and magnetic fields are most generally given by:
\begin{equation}
\label{eq:RealE}
\vec{E}=(E_{dc}+E_0\cos \omega t)\hat{x}+(e_{y}+e_y^r\cos \omega t)\hat{y}+(e_z+e_{z}^r\cos \omega t)\hat{z},
\end{equation}
\begin{equation}
\label{eq:RealB}
\vec{B}=(b_x+f_B b_x^r)\hat{x}+(b_y+f_B b_y^r)\hat{y}+(b_z+f_B B_z)\hat{z}.
\end{equation}
\noindent The component $v_i$ denotes the stray (non-reversing) component of the vector $\vec{V}$ along the i-axis, and $v_i^r$  the reversing $\vec{V}$ component along the same axis. A B-field flip parameter $f_B=\pm 1$ is introduced in (\ref{eq:RealB}). All the field components containing the term $f_B$ reverse with the main magnetic field. Allowing for an ellipticity in the nominally linearly polarized optical field, $\vec{\mathcal{E}}$  becomes:
\begin{equation}
\label{eq:RealEopt}
\vec{\mathcal{E}}=\mathcal{E}(\sin\theta \hat{y}+\cos\theta e^{i\phi} \hat{z}).
\end{equation}
\noindent As discussed in \cite{Tsigutkin2010ParitySystematics}, a rotation operation has to be applied to the fields of (\ref{eq:RealE}), (\ref{eq:RealB}) and (\ref{eq:RealEopt}) so that the rotated $\vec{B}$  is along z. The transition rate  (\ref{eq:R0}), as well as the harmonics amplitudes $R_0^{[1]}$ and $R_0^{[2]}$, acquire then a large number of terms. A series expansion in the field imperfections and $\zeta$ yields a harmonics  ratio $r_0(\theta,f_B)$, in which, in addition to the PV-related term $\zeta/\beta$, terms that transform in the same way as $\zeta/\beta$ under the $\theta$-reversal, are also present. The full expression for $r_0(\theta,f_B)$ in the presence of apparatus imperfections is given in Appendix A. A simplified expression,  that includes the only significantly contributing PV-mimicking term, is the following:

\begin{equation}
\label{eq:r0}
r_0(\theta,f_B)=\frac{4E_{dc}}{E_0}+\left[\frac{4\zeta}{\beta E_0}+\frac{4(b_x+f_Bb^r_x)e_y}{f_BB_zE_0}\right]\cot\theta\cos\phi.
\end{equation}

 There are four different values of $r_0(\theta,f_B)$, corresponding to the two possible values of the polarization angle ($\theta\approx\pm\pi/4$) and magnetic field direction ($f_B=\pm1$), and four different ways to combine these values. These combinations, labeled $K_i$ (i=1,2,3,4), are computed using  the full expression for $r_0(\theta,f_B)$ (see Appendix A) as follows: 

\begin{equation}
\label{eq:Ki_Matrix_equation}
\begin{pmatrix} K_1\\K_2\\K_3\\K_4\end{pmatrix} =
\begin{pmatrix} +1&-1&+1&-1\\-1&-1&+1&+1\\-1&+1&+1&-1\\ +1&+1&+1&+1\end{pmatrix}\cdot
\begin{pmatrix} r_0(\theta_+,+1)\\r_0(\theta_-,+1)\\r_0(\theta_+,-1)\\r_0(\theta_-,-1)\end{pmatrix}.
\end{equation}

 \noindent The values of $K_i$ are given in Table \ref{Table:Ktable}. One of these ($K_1$) yields the ratio $\zeta/\beta$; the others provide important information about parasitic fields and overall measurement consistency. Some of the $K_i$ values are expressed in terms of the polarization parameter $p$, defined as:
\begin{equation}
\label{eq:p_parameter}
p=\cot\theta_+\cos\phi_+-\cot\theta_-\cos\phi_-,
\end{equation}
with $p\approx2$ in the experiment. The angles $\phi_{\pm}$ are the ellipticity-related parameters corresponding to the angles $\theta_{\pm}$. Examination of the terms in $K_1$ shows that a precision determination of $\zeta/\beta$ requires, aside from accurate knowledge of $E_0$, a measurement of the false-PV contribution $e_y b_x^r/B_z$ as well as a measurement of the parameter $p$. Methods to make these measurements are discussed in section IV.

\begin{table}[h]

\caption{The four combinations of harmonics ratio $r_0(\theta,f_B)$ values, corresponding to the two orientations of the polarization angle ($\theta_\pm\approx\pm\pi/4$)  and magnetic field ($f_B=\pm1$). The angles $\phi_\pm$ are the small optical field ellipticity-related parameters for the polarization states with angles $\theta_\pm$, respectively. }

\begin{ruledtabular}
\begin{tabular}{ c c }
\textrm{Combination}&
\textrm{Value}\\
\colrule
$K_1$ & $\left(\dfrac{8\zeta}{\beta E_0}+\dfrac{8b^r_xe_y}{B_zE_0}\right)p$\\[0.4cm]
$K_2$ & $\dfrac{16b_xe_z}{B_zE_0}-\dfrac{32b_y\zeta}{\beta B_zE_0}$\\[0.4cm]
$K_3$ & $-\dfrac{8b_xe_y}{\beta B_zE_0}p$\\[0.4cm]
$K_4$ & $\dfrac{16E_{dc}}{E_0}-\dfrac{16b^r_xe_z}{B_zE_0}+\dfrac{32b_y\zeta}{\beta B_zE_0}$\\[0.2cm]
\end{tabular}
\end{ruledtabular}
\label{Table:Ktable}
\end{table}

\section{\label{sec:level1-Apparatus}Apparatus}
The PV isotopic comparison experiment was carried out with a newly built atomic-beam apparatus which has increased statistical sensitivity and better ability to study and control systematics, compared to that of \cite{Tsigutkin2009ObservationYtterbium,Tsigutkin2010ParitySystematics}.

\begin{figure}
\includegraphics[]{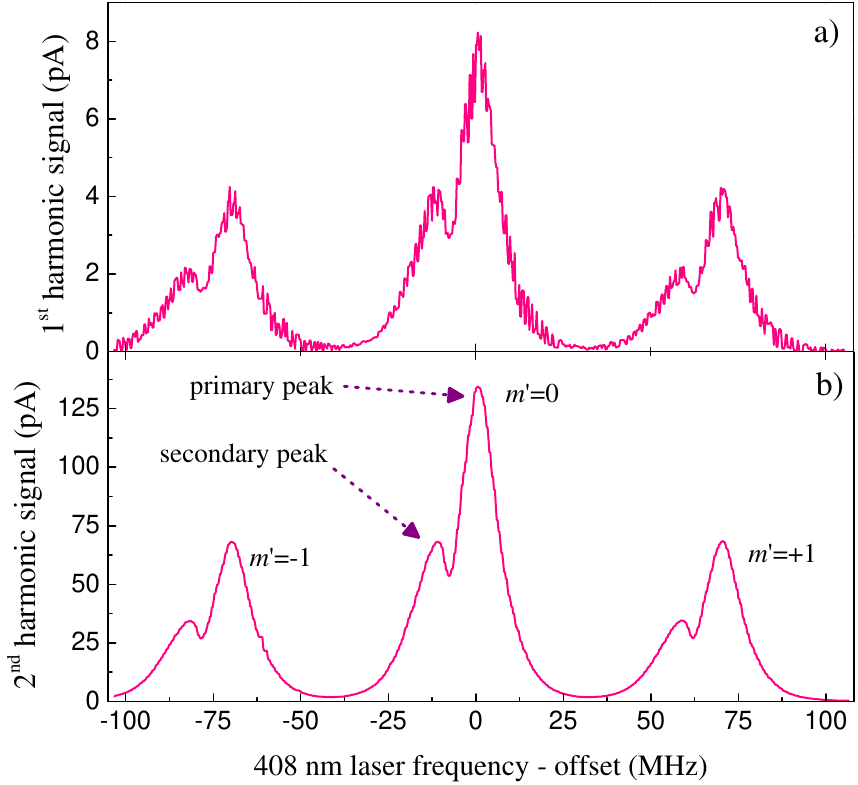}
\caption{(Color online) Spectral profile of the $^{174}$Yb $^1$S$_0$ $\rightarrow$ $^3$D$_1$ transition. a): 1$^{st}$ harmonic present in the excitation rate b): 2$^{nd}$ harmonic. These spectra are obtained by scanning the 408 nm laser frequency around the center of the resonance and measuring the respective harmonic contributions to the photocurrent from the detection region (see text). The three Zeeman components of the transition $^1$S$_0$,  $m=0 \rightarrow$ $^3$D$_1$, $m'=0,\pm1$ are fully resolved in the presence of a $\approx$ 93 G magnetic field in the interaction region. The applied electric field has an ac-amplitude of $E_0$=1000 V/cm and a dc-component of $E_{dc}\approx$14.54 V/cm. The low optical intensity peak is 18 MHz wide (FWHM), and is distorted at higher intensity due to the effects of the off-resonant ac-Stark effect in the presence of the standing-wave field circulating in the PBC (see text). Figure adapted from supplementary material of \cite{Antypas2019IsotopicYtterbium}. }
\label{fig:spectrum}
\end{figure}

A schematic of the in-vacuum setup is shown in fig. 2. An Yb atomic beam is produced with an oven heated to $\approx$ 550 $^{\circ}$C. Atoms exiting the oven nozzle travel a distance of $\approx$28 cm to reach the interaction region, with a mean longitudinal velocity of $\approx$290 m/sec and a transverse velocity spread of $\approx$8 m/s (Full Width at Half Maximum-FWHM). In the interaction region, the atoms intercept the 408 nm standing-wave optical field, tuned to excite the $^1S_0$ $\rightarrow$ $^3D_1$ transition. This light circulates in a power-build-up cavity (PBC), which has a finesse of $\approx$550 and is used to enhance the light power available to excite atoms, but also to suppress the effects of the \textit{M}1-Stark interference. The circulating power is measured by recording the light transmitted through the PBC, and it is actively stabilized, to a level of $\approx$ 55 W. This stabilization results in negligible contribution of intracavity power noise to noise in detection of the excitation rate on the 408 nm transition. The waist ($1/e^2$ intensity radius) of the optical beam in the interaction region is $w_0\approx$ 310  $\mu$m, corresponding to an intensity of $\approx 18$ kW/cm$^2$, or to an optical field applied to the atoms of amplitude $\approx$3.7 kV/cm. This amplitude is about three times greater than the typical amplitude of the quasi-dc field applied in the interaction region $E_0\approx 1.2$ kV/cm.  The intracavity power level is a compromise between the need for large 408 nm excitation rate and unwanted distortion and broadening in the transition lineshape, which appears for an intracavity intensity around 10 kW/cm$^2$ and becomes excessive for intensities above the current level of 18 kW/cm$^2$ . This distortion has been studied extensively in \cite{Stalnaker2006DynamicShapes,Dounas-Frazer2010MeasurementTransitions} and can be removed, if needed, using methods reported in \cite{Antypas2018Lineshape-asymmetryField}. It arises in the presence of an off-resonant ac-Stark effect, induced by the intense  standing-wave field. Owing to the imperfect collimation of the atomic beam, most atoms traversing the standing-wave fly through many nodes and anti-nodes of the field, and in the presence of the ac-Stark effect, experience amplitude, and effectively frequency modulation (the latter occurs due to ac-Stark-induced modulation of the energy levels). This combined amplitude and frequency modulation results in a complex lineshape for the 408 nm transition, that is shown in fig. \ref{fig:spectrum}. 

The required electric field is applied to the atoms with a system of gold-coated electrodes. This system consists of two main plates, approximately $10\times10$ cm$^2$, spaced by 5.5045(20) cm. A set of eight surrounding electrodes is employed to increase field uniformity as well as to apply auxiliary field components in either the y- or z- direction, for systematics studies. Six high-voltage amplifiers and a system of voltage dividers are used to bias the main plates and surrounding electrodes. Simulations of the electric field with COMSOL$^{\textregistered}$ yield a value for the primary field of $[1-2.7(3)\cdot10^{-4}]\cdot V/d$, where \textit{V} is the potential difference between the plates, and \textit{d} is the plate spacing. The non-uniformity of the field within the 1.5 cm wide interaction region (whose diameter is 0.6 mm) is lower than 0.1\%. The magnetic field in the interaction region of 93 G is applied with a pair of round in-vacuum coils, which have nearly Helmholtz geometry. Additional sets of coils are used to cancel the residual field in the interaction region (to within 20 mG), as well as to apply additional field components for studies and control of systematics.

Detection of the 408 nm excitations in the interaction region is done downstream in the path of the atoms using an efficient detection scheme described in detail in \cite{Tsigutkin2010ParitySystematics,Antypas2017TowardsYtterbium}. The fraction of atoms ($\approx 65$\%) that decayed to the $^3$P$_0$ metastable state after undergoing the 408 nm transition, are further excited with $\approx$120 $ \mu$W of 649 nm light to the $^3$S$_1$ state (see fig. \ref{fig:energylevels}), in the region of an optimized light collector (fig. \ref{fig:apparatus}). The light collector directs the induced fluorescence at 556, 649 and 680 nm to a light-pipe which guides light out of the vacuum chamber and onto the surface of a large-area photodiode, whose photocurrent is amplified with a low-noise transimpedance amplifier. This amplifier has a 1 G$\Omega$ transimpedance and $\approx$1.1 kHz bandwidth. The overall detection efficiency of the 408 nm transitions is an estimated 25\% \cite{Tsigutkin2006TowardsYtterbium}.

The 408 nm laser system is a frequency-doubled Ti:Sapphire laser (M$^2$ SolStiS+ECD-X) outputting $\approx$1 W of near-UV light. The laser frequency is stabilized to an internal reference cavity, with a resulting linewidth of less than 100 kHz. The short-term stability of the system is sufficiently good so that we use the internal cavity as the short-term frequency reference. The PBC is stabilized to this reference through frequency-modulation spectroscopy; the PBC length is modulated at 29 kHz using a piezo-transducer onto which one of the cavity mirrors is mounted, and the demodulated PBC transmission is applied back to the piezo with an electronic filter. During an experiment, the laser frequency is locked to the peak of the resonance profile of the atomic transition (see fig. \ref{fig:spectrum}). For this, the Ti:Sapphire frequency is modulated at 138 Hz (with an amplitude of $\approx$200 kHz) and the recorded detection-region fluorescence is demodulated with a lock-in amplifier, whose output is fed back to the laser, through an electronic filter of low ($\approx$1 Hz) bandwidth. This scheme ensures long-term frequency stability for the 408 nm laser system.

The 649 nm laser system, whose output is used to excite the 60 MHz wide $^3$P$_0 \rightarrow ^3$S$_1$ detection transition, is an external-cavity diode laser (Vitawave ECDL-6515R). To suppress frequency noise of this laser, its frequency is locked to the side-of-fringe of an airtight Fabry-Perot (FP) resonator. The resonator length is in turn stabilized with slow feedback to a set laser frequency, whose reading is made with a wavemeter (HighFinesse WSU2). This double-stage scheme ensures short- and long-term stability so that the impact of frequency excursions of the laser on the detection of the 408 nm transition is negligible.

Precise polarization control of the intracavity optical field, as well as continuous measurement of the PBC polarization, are needed in the experiment. The linear polarization of the light coupled to the PBC is set with a half-wave plate mounted on a motorized rotation stage. This polarization is measured with a balanced polarimeter, placed at the output of the PBC. The polarimeter makes use of a Glan-Taylor polarizer that analyzes a small fraction of the light transmitted through the PBC. This light is picked off with a wedge window placed at near-normal incidence in the path of the beam exiting the PBC. The two orthogonal polarization states at the output of the polarizer are measured with a pair of amplified photodetectors. The polarizer axis is set so that the polarimeter is nominally balanced for the $\theta_{\pm}$ polarization angles. The small polarization ellipticity in the PBC, whose value is also required for an accurate PV-effect measurement, is determined using a scheme outlined in section \ref{sec:level3-polarization parameter p}.

Lock-in amplifiers are used to measure the 1$^{st}$ and 2$^{nd}$ harmonics present in the 408 nm excitation rate (models Signal Recovery SR7265 and Zurich Instruments MLFI, respectively). For  typical electric field amplitude $E_0\approx 1$ kV/cm, the contribution to the ratio $r_0$  (\ref{eq:r0}) from the PV effect is $4\zeta/\beta E_0 \approx10^{-4}$. Due to the small size of the 1$^{st}$  harmonic $R_0^{[1]}$, its detection in the presence of a much larger 2$^{nd}$ harmonic amplitude $R_0^{[2]}$ is technically challenging. Two steps are taken to circumvent this issue. First, a field $E_{dc}\approx$ 6.3 V/cm is applied in the interaction region. The resulting contribution $4E_{dc}/E_0$ to the ratio $r_0$ [see (\ref{eq:r0})], of typical value 0.02, is a purely PV-conserving signal, which does not affect the determination of the PV-related effect. The latter is determined through measurements of the change in $r_0$ with polarization angle $\theta$. Second, the signal directed to the lock-in measuring $R_0^{[1]}$ is filtered with an amplified band-pass filter, which provides a gain of 101.67(22) for the 1$^{st}$ harmonic while attenuating the 2$^{nd}$ harmonic $\approx$50 times. These two steps result in $R_0^{[1]}$ and $R_0^{[2]}$ signals of comparable size presented to the respective lock-in amplifiers. Finally, to avoid potential systematic effects due to the changing signal levels when measuring different isotopes, a variable-gain amplifier is used to adjust the signal level at the output of the detection-region photodetector. The gain values in this amplifier are related to the different isotopic abundances of the four Yb isotopes measured, such that the same signal level is always presented to the lock-ins, regardless of isotope measured.

\section{\label{sec:level1-Investigation-Systematics}INVESTIGATION OF SYSTEMATIC effects AND RELATED ERRORS}
In this section we present a detailed analysis of systematic contributions and uncertainties related to the isotopic comparison measurements. These uncertainties are either due to the limited accuracy of the various calibrations or imperfect estimates of the contribution of PV-mimicking effects. We begin by discussing the various PV-data calibrations and the errors in these, since the latter dominate the total systematic uncertainty in the present experiment. We then present an analysis of false-PV contributions and the related uncertainties. Finally, auxiliary experiments done to ensure consistency with our model of harmonics ratios, as well as to investigate potentially unaccounted-for systematics, are discussed at the end of the section.

\subsection{\label{sec:level2-Calibrations of PV data}Calibrations to PV-data and related uncertainties}

\subsubsection{\label{sec:level3-408 nm transition saturation}408 nm transition saturation}

In the absence of saturation in the Stark-induced transition, the 408 nm signal grows as $E^2$. In the present experiment the transition is weakly saturated. This slight saturation affects the measurement of the harmonics ratio $r_0$, and a correction needs to be made. The transition rate can be generally expressed as \cite{Wood1999PrecisionCesium}:
\begin{equation}
\label{eq:RateWithSaturation}
R=\frac{kE^{2}}{1+\frac{E^{2}}{E_s^2}}.
\end{equation}
\noindent The parameter \textit{k} is an overall constant (which depends on the light power in the PBC), $E_s$ the saturation electric field, and $E=E_0\cos\omega t+\zeta/\beta$ includes the applied electric field and the effective electric field $\zeta/\beta$ that results from the PV ($|\zeta/\beta| <<E_0$). The field $E_s$ depends on the intensity  of the 408 nm light exciting atoms. The rate \textit{R} is saturated when $E_0$ becomes comparable to $E_s$. In the present experiment, the 408 nm transition in the atomic beam is weakly saturated ($E_0/E_s\approx 0.1$). To quantify the impact on the harmonics ratio, we expand \textit{R} in terms of the parameter $(E_0/E_s)^2$, and compute $r_0$. To first order in this parameter, the modified ratio is:
\begin{equation}
\label{eq:r0WithSaturation}
r_0=\frac{\zeta}{\beta E_0}\bigg(1-\frac{1}{2}\frac{E_0^2}{E_s^2}\bigg).
\end{equation}
\noindent PV data need to be therefore divided by:
\begin{equation}
\label{eq:CsFactor}
C_s=1-\frac{1}{2}\frac{E_0^2}{E_s^2}.
\end{equation}
\noindent Similar analysis shows that the $2^{nd}$ harmonic in the transition rate is also diminished in the presence of saturation, by a factor $(1-E_0^2/E_s^2)$.

In the presence of transition saturation, harmonics higher than the $2^{nd}$ emerge in the rate of eq. (\ref{eq:RateWithSaturation}). We make use of a $4^{th}$ harmonic amplitude to measure the saturation parameter $E_s$. The ratio of $4^{th}$ to $2^{nd}$ harmonic amplitudes (to first order in $(E_0/E_s)^2$) is given by $E_0^2/4E_s^2$. Measurements of this ratio with varying $E_0$ (in the range 1-2.5 kV/cm) are made to determine $E_s$.

\begin{figure}
\includegraphics[]{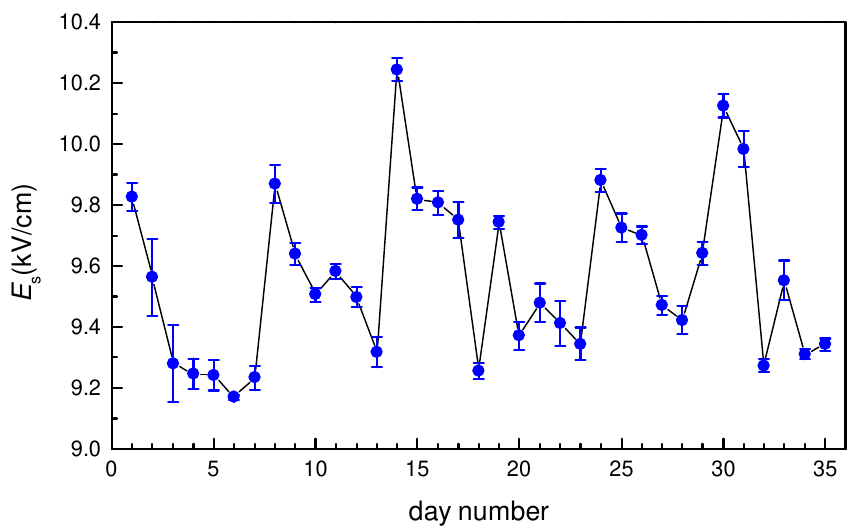}
\caption{(Color online) Measurements of the saturation electric field for the $0 \rightarrow 0$ component of the 408 nm transition, made on each of the 34 days in which PV-data where acquired. Each data point represents the average of four measurments, two of which were made for angle $\theta_+=+\pi/4$ and another two for $\theta_-=-\pi/4$.}
\label{fig:Esat}
\end{figure}

We show in fig. \ref{fig:Esat} measurements of the parameter $E_s$, made in each of the 34 days in which actual isotopic comparison PV-data were acquired. A periodic pattern can be observed in the data that involves a gradual decrease in $E_s$, followed by a recovery. This effect is currently not fully understood; however, as we observe, it is generally correlated with gradual deterioration of the in-vacuum PBC mirrors, in the presence of the intense near-UV light. Typically, operation of the PBC for a few days results in a decrease in the cavity finesse and power buildup of about 30\%. The gradual decrease in $E_s$ should be occurring due to an increase in the intra-cavity circulating power (which corresponds to an increase in the degree of saturation in the transition rate). Since the power transmitted through the PBC is actively stabilized, the observed effect implies that the transmission of the cavity output coupler gradually decreases.  Recovery of the cavity mirrors is possible by exposing them to partial atmosphere (tens of mbar) for $\approx$1 min, in the presence of the intense 408 nm light. The recovery process generally results in an increase of $E_s$. As seen in fig. \ref{fig:Esat}), the saturation field $E_s$ increases following venting of vacuum system which was done to recover PBC mirror performance before days \# 1, 8, 14, 19, 24, 30. We assume an error of 3\% in the daily $E_s$ value, to take into account possible drifts of this parameter over the 8-16 hr long PV-run.

\subsubsection{\label{sec:level3-polarization parameter p}Polarization parameter \textit{p}}

The 408 nm polarization parameter $p$ of eq. (\ref{eq:p_parameter}) needs to be precisely measured for an accurate $\zeta/\beta$ determination. For angles in the range $\vert\theta_{\pm}\vert=\pi/4\pm0.02$ and $\vert\phi_{\pm}\vert\leq 0.06$, this parameter can be approximated (with an error of a few parts per 10$^5$) as $p\approx p_{\theta} \cdot p_{\phi}$, with:
\begin{equation}
\label{eq:pTheta}
p_\theta=\cot\theta_+-\cot\theta_-,
\end{equation}
\begin{equation}
\label{eq:Phi}
p_\phi=\cos\phi_+-\cos\phi_-.
\end{equation}
\noindent This separation of variables simplifies the determination of $p$. In the following, we discuss how $p_\theta$ and $p_\phi$ are measured. 

Continuous measurements of $\theta$ during PV data acquisition are made with the PBC polarimeter described in section \ref{sec:level1-Apparatus}. Prior to commencing an acquisition run, a calibration of the PBC polarimeter is required. To perform this calibration, measurements of the relative sizes of the three transition components in the 408 nm spectrum (see fig. \ref{fig:spectrum}b) are used to read the intracavity light polarization angles $\theta_{\pm}$ (nominally $\pm\pi/4$); these angles are correlated with the concurrent readings the PBC polarimeter, thereby providing a calibration of the polarimeter. Subsequent measurements of the light transmitted through the PBC during a many-hour-long PV run provide an accurate tracking of the angles $\theta_{\pm}$. A detailed description of the method to determine the initial $\theta_{\pm}$ angles using the atoms as polarization probes, including the effects of apparatus imperfections, is given in Appendix \ref{sec:level1-AppendixB}.

\begin{figure}
\includegraphics[]{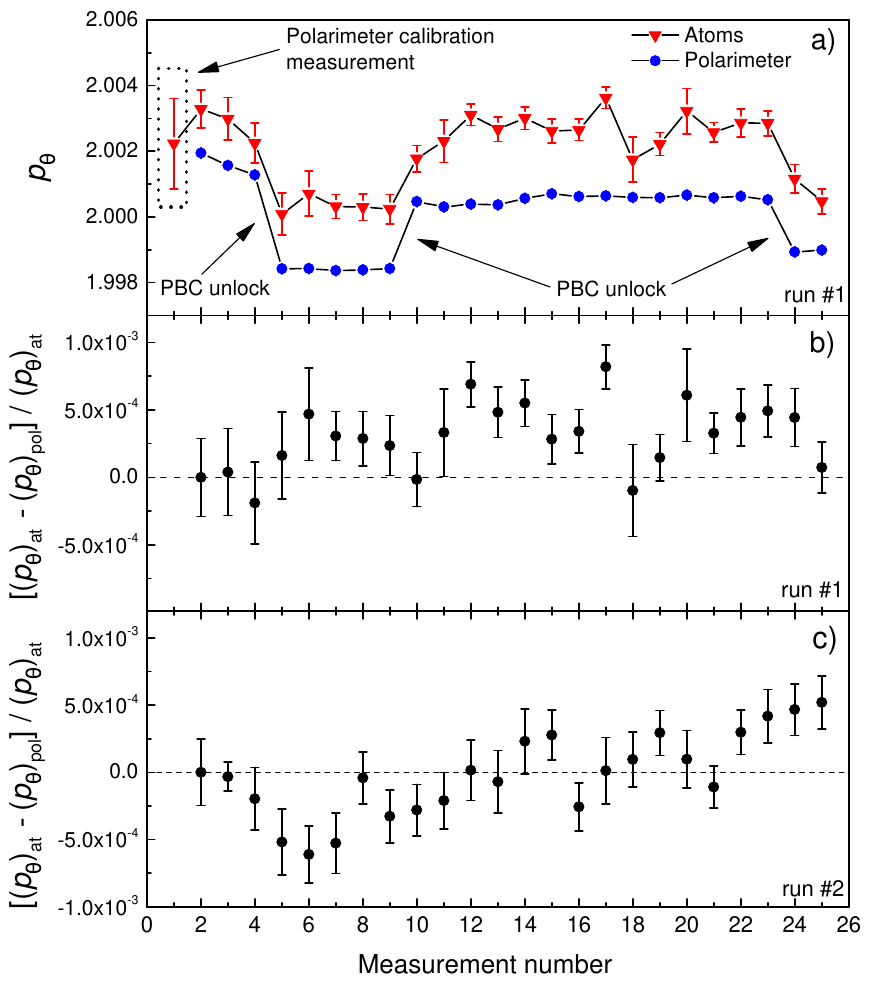}
\caption{(Color online) a) Comparison of $p_{\theta}$ measurements made with atoms and those made with the PBC polarimeter, over a 12 hr period. Error bars (smaller than data points for the polarimeter data) represent standard errors of the corresponding mean values. The first data point in the plot represents the initial reading of angles $\theta_{\pm}$ using the 408 nm spectrum. These readings are used to assign an initial value to the corresponding polarimeter readings. The statistical error in this first $p_{\theta}$ measurement explains the relative offset between the 'atoms' and 'polarimeter' points in the second measurement (points with measurement \#2). The polarimeter calibration measurement (measurement \#1) has greater error than subsequent measurements, as it has smaller integration time compared to the time devoted to measure subsequent points.  The sub-0.1\% statistical error of this calibration measurement is negligible compared to the $\sim$ 1\% statistical error of a daily block of PV data. The PBC was unlocked several times, with the duration of each pause in the range 5-10 min. Shorter ($\approx$10 s) interruptions in the PBC lock were also made, and have no visible impact on polarization. b) Relative difference in $p_{\theta}$ readings between the two methods for the data shown in a). The  offset between the 'atoms' and 'polarimeter' values at the start of the run ( see points with measurement \#2 in plot a)), is of statistical nature, and is removed in b), to allow for a study of relative drifts between the two determinations. c) Results of $p_{\theta}$ differences measured in another 12 hr-long run.}
\label{fig:Polarization theta}
\end{figure}

The uncertainty in $p_{\theta}$ has two contributions: the statistical uncertainty associated with the initial $\theta_{\pm}$ measurement using the atoms, and the systematic uncertainty arising from drifts in the readings of the polarimeter at the output of the PBC over a many-hour period. The statistical uncertainty (typically $<$0.1\% of the PV effect) is added in quadrature with the statistical error in a block of data acquired in a daily run. To make an estimate for the systematic uncertainty, we took two long sets of polarization data. In these runs, following the initial correlation of the $\theta_{\pm}$ readings with the polarimeter readings, the $p_{\theta}$ measurements made with the two methods were compared over a period of 12 hours. These data are presented in fig. \ref{fig:Polarization theta}. During these runs, the PBC was unlocked several times, to investigate the effect of thermal cycles of the PBC optics on the actual polarization angle (read with the atoms), and well as on its measurement with the polarimeter. The data show that unlocking the PBC for minute-long periods of time, does have an impact on the intra-cavity polarization angle (fig.  \ref{fig:Polarization theta}a). These polarization shifts are nevertheless tracked well by the polarimeter, as seen in fig. \ref{fig:Polarization theta}b. The relative drifts between the $p_{\theta}$ determinations made using the 408 nm resonance profile and those  made using the polarimeter are always less than $10^{-3}$ of the nominal value $p_{\theta}=2$. We assign a 10$^{-3}$ fractional systematic uncertainty in determining $p_{\theta}$.

The light ellipticity-related parameter $p_{\phi}$ is determined through measurements made using signals from the atoms. The idea is to observe a term in the harmonics ratio of the $m'=\pm 1$ components of the 408 nm transition, that has a dependence on the angle $\phi$. Expressions for the excitation rate for these components, as well as the corresponding harmonics ratios $r_{+1}$ and $r_{-1}$ in the presence of field imperfections, are given in Appendix A. The difference $r_{+1}-r_{-1}$ (retaining terms up to $2^{nd}$ order in the various field imperfections), is given by:

\begin{equation}
\label{eq:Measurephiequation}
r_{+1}-r_{−1}=\frac{8e_z}{E_0} \tan\theta\sin\phi.
\end{equation}

\begin{figure}
\includegraphics[]{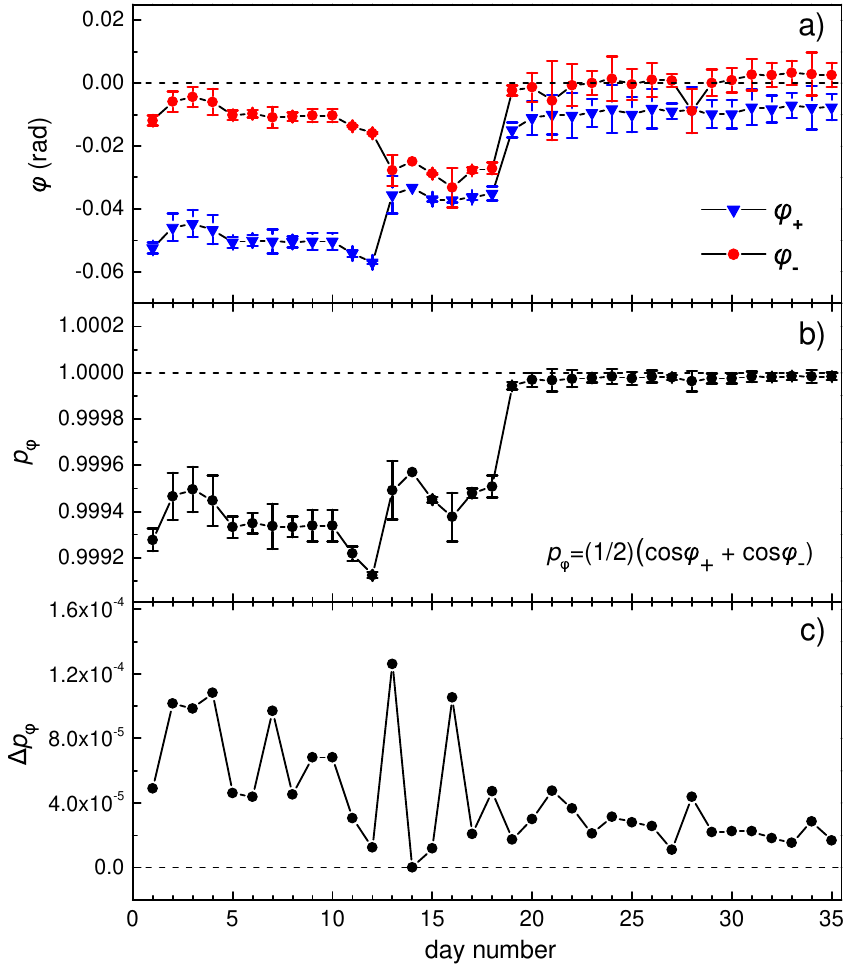}
\caption{ a) (Color online) Angles $\phi_{+}$ and $\phi_{-}$ determined on each of the 34 days of actual PV-data acquisition. Each data point represents the mean value of two measurements, of which the first was done before the start of the PV-run, and the second after the end of the run. The error bars correspond to the standard deviation of the mean. The changes in these values on days \#13 and \#18 are due to a readjustment of the tilt of the half-wave plate that controls the 408 nm light polarization coupled to the PBC. b) Corresponding $p_{\phi}$ parameter for the values shown in a). Error bars come from propagation of errors in the data points of a). c) Plot of the error-bar values for the data shown in b). The maximum error is $1.2\cdot 10^{-4}$.}
\label{fig:ellipticity}
\end{figure}

A measurement of  $r_{+1}$ and $r_{-1}$ with an enhanced field component $e_z$, ($e_z/E_0\approx\pm0.06$) allows for extraction of $\phi_+$ or $\phi_-$, corresponding to polarization states with $\theta_+$ or $\theta_-$ respectively. The overall accuracy is determined by statistics (the ratio $e_z/E_0$ is known to within 1\%, and $\tan\theta$ is measured with sub-1\% uncertainty).

Measurements of $\phi_{\pm}$ were made before the start and after the end of each of the many-hour-long PV-runs, employing both orientations of the magnetic field $B_z$. The $\phi_{\pm}$ value for the corresponding daily block of PV-data is taken as the mean of the initial and final measurements, and the error assigned to this mean is the standard deviation of the two values. We show the results of these measurements as well as resulting parameter $p_{\phi}$ and the error in its determination in fig. \ref{fig:ellipticity}. We assign a fractional error of $1.2\cdot 10^{-4}$ in the determination of $p_{\phi}$. This is a negligible contribution to the overall error in the polarization parameter \textit{p}, an error dominated by the $10^{-3}$ fractional error in $p_{\theta}$.

\subsubsection{\label{sec:level3-effect of partial peak overlap}Effect of partial peak overlap}
The applied magnetic field in the interaction region results in a resolved spectrum for the $^1$S$_0 \rightarrow ^3$D$_1$ transition (fig \ref{fig:spectrum}). A small residual overlap between the different peaks is still present, however, and its effect on the PV-measurements needs to be considered. In the presence of the overlap, the transition rate at the spectral peak position of the $0\rightarrow0$ transition component (where the PV-data are acquired) is given by:

\begin{equation}
\label{eq:Rpeak_overlap}
R_0^{'}=R_0+h\cdot R_{-1}+h\cdot R_{+1}.
\end{equation}

\noindent The terms $R_{-1}$ and $R_{+1}$ are the rates of the $0\rightarrow -1$ and $0\rightarrow +1$ components, and \textit{h} is a parameter that quantifies the contribution of the wing of a peak to the signal of the adjacent peak, measured to be $h=4.2(4)\cdot10^{-4}$. In formulating the total rate $R_0^{'}$ in eq. (\ref{eq:Rpeak_overlap}), quantum interference between the transition amplitudes of the different Zeeman sub-levels is not considered. Such an effect does not take place in our system, since the emitted fluorescence light from de-excitation of atoms has different polarizations for the three excited-state sub-levels. Because of this, the corresponding excitation paths ($m=0\rightarrow m'=0,\pm1)$ can be distinguished and amplitude interference does not occur. The resulting harmonics ratio $r^{'}_0$ can be computed from (\ref{eq:Rpeak_overlap}), and from that, the corresponding combination $K^{'}_1$ (Table \ref{Table:Ktable}), can be determined:

\begin{equation}
\label{eq:K1WithSaturation}
K^{'}_1=\frac{16\zeta}{\beta}(1-2h).
\end{equation}

To derive (\ref{eq:K1WithSaturation}), field imperfections, which generally have a greater impact on the PV-measurements in the $0\rightarrow  \pm1$ transitions, were neglected. This is a reasonable simplification. As discussed in section \ref{sec:level3-Other consistency checks}, the PV-effect on the $0\rightarrow \pm1$ transitions, is, to within 2\%, consistent in magnitude with the effect measured on the $0\rightarrow 0$ transition. An additional 2\% correction to the small calibration parameter of order $h$, would have a negligible impact on the PV-measurements taken on the $0\rightarrow 0$ component. The negative sign in the signal contribution from the $0\rightarrow  \pm1$ transitions is expected, since the PV-effect for these transitions is of opposite sign compared to that of the $0\rightarrow 0$ component (see Appendix A). To correct for the effect of the patial overlap of the different Zeeman components in the $^1$S$_0 \rightarrow ^3$D$_1$ transition, the PV-measurements are divided by a factor $C_{overlap}= (1-2h)$.

\subsubsection{\label{sec:level3-Transit time from interaction to detection region}Transit time from interaction to detection region}

Due to the time required for excited atoms to reach the detection region and be measured, there is a phase delay in the detected excitation rate, relative to the electric field phase in the interaction region. This would not be an issue for a  beam of atoms all moving with the same longitudinal velocity; however, because of the longitudinal velocity spread in the atomic beam, atoms in different velocity classes are detected at different times. This leads to a slight mixing of phases in the measured rate for these different classes, and to a frequency-dependent attenuation of the amplitude of each harmonic. The result of this attenuation is a detected harmonics ratio $r_0$ that is slightly larger than the actual one. This effect was modeled in \cite{Tsigutkin2010ParitySystematics}. We correct for it by dividing the measured $r_0$ by a factor $C_{transit}$=1.00285(10). This factor is an order of magnitude lower than that in \cite{Tsigutkin2010ParitySystematics}. The reduction is due to the lower electric-field frequency (19.9 Hz) in the present experiment, compared to that of the previous one (76 Hz). The assigned error in $C_{transit}$ comes from the assumed uncertainty in the temperature of the Yb oven ($\pm$ 50 $^{\circ}$C) and from the assumed 0.5 cm uncertainty in the distance between the interaction and detection regions. The expected phase-delays in 1$^{st}$ and 2$^{nd}$ harmonic signals present in the transition rate (- 4.8$^{\circ}$ and - 9.6$^{\circ}$) are detected correctly, to within 0.5$^{\circ}$. A 0.5$^{\circ}$ error in the detected phase of a given harmonic in the excitation rate, would result in a fractional decrease of $5\cdot10^{-5}$ in the measured harmonic amplitude. The uncertainties in the measured PV effect arising from such  small phase uncertainties in detecting the $1^{st}$ and $2^{nd}$ harmonics, are negligible. 

\subsubsection{\label{sec:level3-photodetector response calibration}Photodetector response calibration}
The detection-region photodetector (PD) has a finite bandwidth, measured to be 1.1 kHz. The PD low-pass-filter behavior at the 1$^{st}$- and 2$^{nd}$- harmonic frequencies present in the transition rate (19.9 Hz and 39.8 Hz, respectively) is expected  to have an impact on the measured  ratio $r_0$. To quantify this impact, we measured the frequency-dependent response of the PD, relative to that of a fast photodetector (Thorlabs PDA100, 220 kHz bandwidth). Using a light-emitting-diode as a source of sinusoidally modulated light, we measured with the PD a ratio of amplitudes at 39.8 Hz and 19.9 Hz, which was 1.00040(17) times greater than the ratio determined with the fast detector. The error in the measured amplitudes ratio is mainly statistical. The measured $r_0$ values are scaled down by $C_{PD}$=1.00040(17) to compensate for the PD finite response time.

\subsubsection{\label{sec:level3-PD signal conditioning}PD signal conditioning calibration}

There is an overall calibration factor $C_e$ relating the harmonics-ratio value recorded in the laboratory PC to the actual ratio at the output of the PD. This factor needs to be precisely measured. As part of the effort to improve detection conditions for the small 1$^{st}$-harmonic signal in the transition rate, the PD signal is bandpass-amplified and then measured with a lock-in amplifier (see section \ref{sec:level1-Apparatus}). The 1$^{st}$-harmonic reading is recorded in the computer, as is the reading from another lock-in that measures the 2$^{nd}$ harmonic directly at the PD output. The calibration factor $C_e$ was measured by replacing the PD with an electronic circuit that adds two known signals at the $\omega$ and 2$\omega$ frequencies. This circuit attenuates the $\omega$ signal to simulate the amplitude level in the actual experiment. The transfer function of this circuit for each of the two signal paths was measured at the $10^{-4}$ level. The inputs to the circuit come from a dual-channel function generator (Keysight 33510B) and are measured with a laboratory multimeter (Keysight 34410A), whose measurements agree with those made with an identical unit, at the $10^{-4}$ level. A comparison of the known harmonics ratio at the output of the adder-circuit, to the reading in the computer, determines $C_e$.

Many different measurements of $C_e$ where made, with varying signal sizes as well as phase-delays between the lock-in reference phases and the corresponding detected phases. These measurements were carried out twice: before the start of the PV-data acquisition campaign, and after its end. The first measurement yielded a value $C_e$=101.52(5) and a second a value $C_e$=101.82(1). We assign the value of 101.67(22) to $C_e$, which is the mean of the two results. The 0.22\% error in $C_e$ is the standard deviation of the two measurements.

This inadvertent drift in the $C_e$ calibration gives rise to the main systematic uncertainty in this experiment. Since the PV-data were acquired in a pattern that involved alternating measurements between isotopes, however, the impact of this drift on the actual isotopic comparison should be minimal.

\subsubsection{\label{sec:level3-electric field calibration}Electric-field calibration}

Accurate knowledge of the electric field applied to the atoms is needed to relate a determination of $K_1$ (see Table \ref{Table:Ktable}) to the ratio $\zeta/\beta$. There are two dominant uncertainties in the electric field. The first is an uncertainty in the calibration of the voltage monitor outputs in the two high-voltage amplifiers (model TREK 609B), used to apply voltage to the main field plates. The corresponding error in the applied voltage is a fractional $6\cdot10^{-4}$. The second uncertainty comes from imperfections in the construction of the field-plate system and the finite accuracy in measuring the field-plate spacing. This spacing was measured at several different places with a precision micrometer. The variation in the mean spacing (5.5045 cm) was found to be 0.002 cm, which corresponds to a fractional uncertainty in the spacing of $\approx4\cdot10^{-4}$, and to the same contribution to the overall electric field error.

\subsection{\label{sec:level2-False PV signals}False-PV signals and related uncertainties}

In this subsection we discuss the methods to study and control known systematic contributions to the measurements which mimic the PV effect.

\subsubsection{\label{sec:level3-eybx contribution}$e_yb_x^r$ contribution}

Examination of the combination $K_1$ (Table \ref{Table:Ktable}), shows that the coupling of a stray $e_y$ field to the reversing magnetic field component $b_x^r$  gives rise to the false-PV contribution proportional to $e_yb_x^r/B_z$, which directly competes with $\zeta/\beta$. The strategy to handle this contribution is to minimize $b_x^r$, and then measure the residual effect periodically during the PV data acquisition, and, if needed, apply a correction to the PV-data.

\begin{figure}
\includegraphics[]{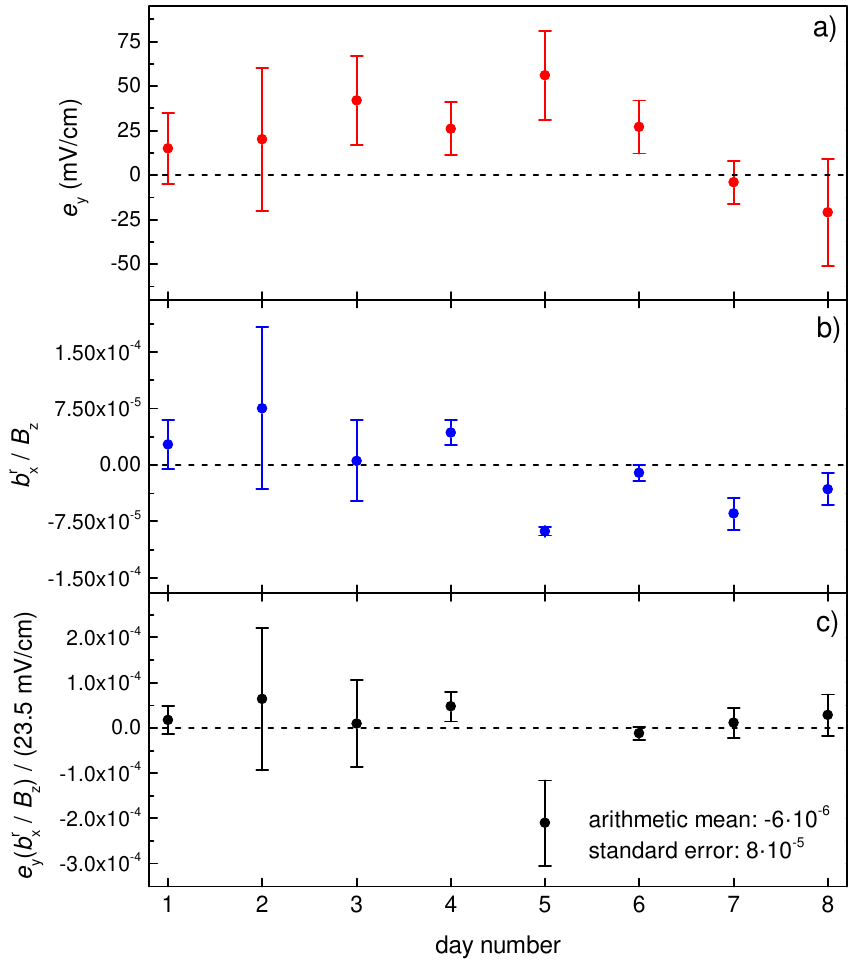}
\caption{(Color online) a) Measurements of the stray field $e_y$, made regularly during the isotopic comparison PV-run. This field is always 75 mV/cm or less in magnitude. b) Measured residual $b^r_x/B_z$ values. The compensated ratio is  stable at the $10^{-4}$ level. c) Fractional contribution of the $e_yb^r_x/B_z$ term to the measurement of the PV effect.}
\label{fig:bacxey}
\end{figure}

To measure $b_x^r$ we apply an enhanced $e_y$ field, with $e_y/E_0=\pm0.062(1)$, and observe the change in $K_1$ as the polarity of this enhanced field is reversed. This allows one to isolate the $e_yb_x^r/B_z$ term and measure $b_x^r/B_z$. The typical value for this misalignment is $1-2\cdot 10^{-3}$. We then use shimming coils to to apply a reversing field to null $b^r_x$. With this procedure the residual $b_x^r/B_z$ ratio is measured to be $10^{-4}$ or less. We find that this cancellation is very stable with time (over month-long periods). Readjustment is only required when the alignment of the PBC optical axis (that defines the x-axis in the coordinate system) is changed. Such a change was only made once during the isotopic comparison data run.

With a suppressed $b_x^r/B_z$ ratio, one has to  monitor $e_y$ during the PV-data campaign. To measure $e_y$ we make use of the combination $K_3=-16e_yb_x/E_0B_z$. Another set of coils is used to apply an enhanced $b_x$, with $b_x/B_z\approx \pm0.0390(6)$. Observation of the variation of $K_3$ with a sign flip in $b_x/B_z$ is used to determine $e_y$.

We show measurements of $e_y$ and the residual $b_x^r/B_z$ ratio in fig. \ref{fig:bacxey}. These measurements were made concurrently, at regular intervals during the isotopic comparison PV run. The term $e_yb_x^r/B_z$ was never greater than $2\cdot 10^{-4}$ of the measured PV-effect. The (arithmetic) mean value of the systematic is smaller than $10^{-5}$. The error (standard error of the arithmetic mean) is less than $10^{-4}$ of the PV effect. We conclude that the contribution of this systematic to the PV measurements is negligible. We did not make use of weights in this statistical analysis, since the results of fig. \ref{fig:bacxey} come from short acquisition runs, therefore the corresponding error bars may not represent errors accurately.

\subsubsection{\label{sec:level3-dk1dbx}d$K_1$/d$b_x$ systematic}

The $e_yb^r_x/B_z$ term is the only parasitic contribution in $K_1$, within our model for the harmonics ratio, and up to $2^{nd}$ order in field imperfections. During auxiliary experiments that involved consecutive application of all possible field imperfections to the atoms, as a check for unaccounted-for systematic contributions, we discovered a dependence of $K_1$ on the non-reversing $b_x$ component of the magnetic field. $K_1$ changes with $b_x$ at a rate of $\approx$ −3\%/G, for the $B_z$=93 G leading field. The origin of this effect is currently not understood. We did investigate its dependence on other parameters. No dependence was found on applied non-reversing or reversing electric-field components, or the amplitude of the leading electric field $E_0$. We did observe a $\cot\theta$ dependence of the effect, like the PV effect itself has.

\begin{figure}
\includegraphics[width=8.7cm]{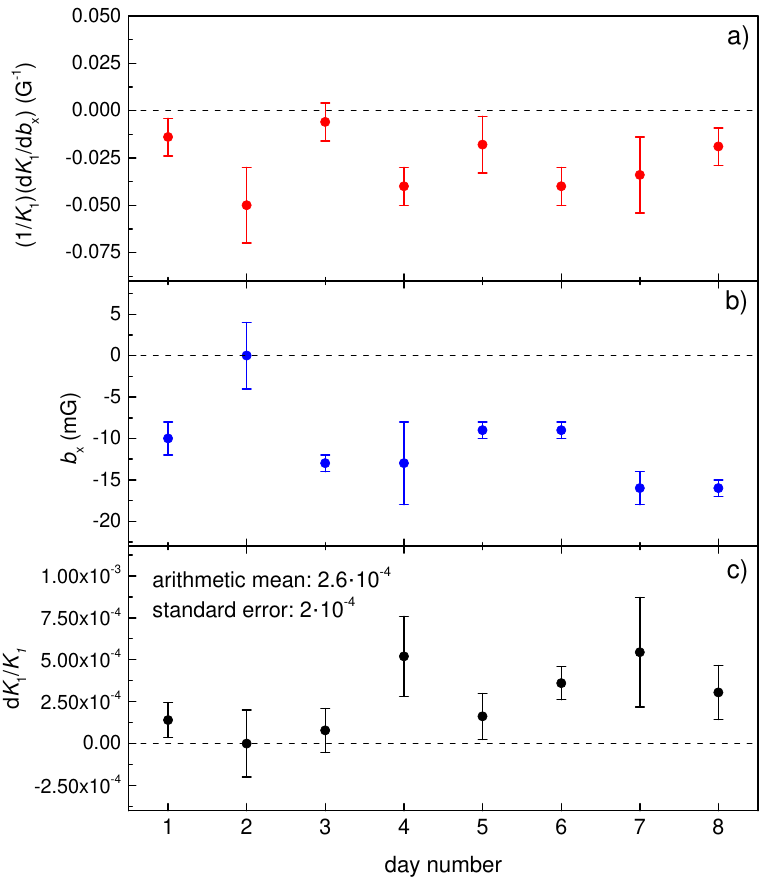} 
\caption{(Color online) a) Fractional change in \textit{K}1 per G of the applied field $b_x$. b) Measured residual $b_x$ field. c) Fractional change in \textit{K}1, inferred from the data of a) and b).}    
\label{fig:dk1-k1}
\end{figure}

This spurious effect is periodically measured and corrections to the PV-data are made. To measure the residual $b_x$ field, we make use of $K_3\approx-16e_yb_x/B_zE_0$, in a manner similar to that described earlier for the measurements of the $e_y$ field. Here we apply a known $e_y$, so that $e_y/E_0\approx\pm0.0644(10)$, and observe the change in $K_3$, correlated with a polarity flip in $e_y$. In measurements made periodically during the PV-data run campaign, the observed $b_x$ values were always smaller than 20 mG. 

We show the measured dependence of $K_1$ on $b_x$  in fig. \ref{fig:dk1-k1}a, along with calibration measurements of the spurious effect in fig \ref{fig:dk1-k1}b. These data were taken regularly during the PV-data acquisition. The corresponding fractional $K_1$ change, inferred from the data of a) and b), is shown in fig. \ref{fig:dk1-k1}c. The arithmetic mean value of the change is $2.6\cdot10^{-4}$. We subtract this fraction from all PV data to account for this systematic effect, and assign a fractional uncertainty $2\cdot10^{-4}$, which represents the standard error of the arithmetic mean value. As in the studies related to the $e_yb^r_x/B_z$ systematic, use of weights in the statistical analysis is avoided.

\subsubsection{\label{sec:level3-Edc and saturation}$E_{dc}$ and transition saturation}

The $E_{\mathrm{dc}}$ field, applied to improve conditions in the Stark-PV interference detection, gives rise to a false-PV signal in the presence of saturation in the transition. To illustrate this, we consider the harmonics ratio of eq. (\ref{eq:r0ideal}):

\begin{equation}
\label{eq:r0EffectofEdc}
r_0(\theta,f_B)=C_s\big(\frac{4E_{dc}}{E_0}+\frac{4\zeta}{\beta E_0}\cot\theta \big),
\end{equation}

\noindent where $C_s$ represents the slight saturation-related reduction in $r_0$ and is given by (\ref{eq:CsFactor}). This factor depends on the overall excitation rate. The corresponding combination $K_1$ is given by:

\begin{equation}
\label{eq:KiEdcwithSaturation}
K_1=\frac{16E_{dc}}{E_0}(C_{s+}-C_{s-})+\frac{8\zeta}{\beta E_0}(\cot\theta_{+}C_{s+}-\cot\theta_{-}C_{s-}).
\end{equation}

\noindent The saturation factors $C_{s+}$ and $C_{s-}$ correspond to the two angles $\theta_+$ and $\theta_-$, respectively, and are generally slightly different, due to unequal excitation rates for the angles $\theta_+$ and $\theta_-$. Unequal excitation rates occur because $\theta$ is not precisely set to either $+\pi/4$ or $-\pi/4$. The quantity in parenthesis in the second term of (19) can be approximated as:
\begin{equation*}
\label{}
\frac{1}{2}(C_{s+}+C_{s-})(\cot\theta_+-\cot\theta_-),
\end{equation*}
\noindent with an accuracy at the $10^{-5}$ level for angles $\theta_\pm
\approx\pm\pi/4$ and the typical value $\approx 0.01$  for $C_{s+}$ and $C_{s-}$. The parameter $K_1$, from which the PV-related parameter $\zeta/\beta$ is extracted, can be then expressed as:

\begin{multline}
\label{K1EdcWithSaturationApprox}
K_1=\frac{16E_{dc}}{E_0}(C_{s+}-C_{s-})\\
+\frac{1}{2}(C_{s+}+C_{s-})\frac{8\zeta}{\beta E_0}(\cot\theta_+-\cot\theta_-).
\end{multline}

\noindent Since $|C_{s+}-C_{s-}|\neq 0$, extraction of $\zeta/\beta$ from $K_1$ is influenced by the presence of the first term in (\ref{K1EdcWithSaturationApprox}). This term is linear in $E_{dc}$ and leads to a fractional false-PV systematic:  
\begin{equation}
\label{EdcWithSatFractSyst}
\frac{E_{dc}}{\zeta/\beta}|C_{s+}-C_{s-}|.
\end{equation}
During a PV-run we observe excitation rates for the two polarization angles $\theta_{\pm}$ which typically differ by $\approx$ 0.5\%. This is because these angles are not precisely $\pm\pi/4$. Given that the saturation electric field $E_s$ (\ref{eq:CsFactor}) grows as the square root of the signal, we can estimate that for the typical $E_0$=1 kV/cm and $E_s$=10 kV/cm, the quantity $|C_{s+}-C_{s-}|\approx 2.5\cdot 10^{-5}$. Using the value of $\vert E_{dc}\vert$=6.3 V/cm of this experiment, and the measured $\vert \zeta/\beta \vert\approx$23.5 mV/cm, we find that the false-PV term of (\ref{EdcWithSatFractSyst}) is a fractional 0.7\%.

We handle this systematic by averaging PV-data taken with opposite $E_{dc}$ polarities. The more precisely $E_{dc}$ is reversed, the better the suppression of the related systematic. A good reversal is achieved with feedback on the $E_{dc}$ value. To implement this, we make use of the combination $K_4$ (see table \ref{Table:Ktable}), which, to an excellent approximation, is equal to $16E_{dc}/E_0$ (other terms in $K_4$ are suppressed by at least $10^4$ times relative to this term). While data are being acquired, $K_4$ is monitored. Every time the $E_{dc}$ polarity is flipped to negative, an adjustment is made to the new $E_{dc}$ setting, to correct for small differences between the magnitudes of the previous two $K_4$ measurements, one of which  corresponds to $E_{dc}>0$ and the other to $E_{dc}<0$. As a result, the total static field along x (i.e. the sum of the $\vert E_{dc}\vert\approx$6.3 V/cm field and a stray field) is reversed to within 5-10 mV/cm, leading to a practically complete suppression of the related systematic effect.

We note that the slight mismatch between the saturation-related parameters $C_{s+}$ and  $C_{s-}$ does not affect the determination of the calibration factor $(1/2)( C_{s+}+ C_{s-})$ in (\ref{K1EdcWithSaturationApprox}).  This factor is determined as the average of measurements of the parameter $C_s$ made on both angles $\theta_+$ and $\theta_-$.

\subsection{\label{sec:level2-zetabeta sign and consistency checks}$\zeta/\beta$ sign and consistency checks}

In this sub-section we discuss observations made to establish the sign of $\zeta/\beta$. The present determination disagrees with that of the 2009 experiment \cite{Tsigutkin2009ObservationYtterbium,Tsigutkin2010ParitySystematics},  which we have traced to a sign error in the analysis code employed in that work. We also provide the results of auxiliary experiments done to ensure  consistency between measurements and our model for the expected PV effect. 
\subsubsection{\label{sec:level3-zetabeta sign determination}$\zeta/\beta$ sign determination}

The primary method to determine the sign of $\zeta/\beta$  is to study the sign of the  term ($\zeta/\beta)\cot\theta$ in the harmonics ratio of (\ref{eq:r0}), in relation to the signs of other terms in this ratio. The latter signs are unambiguously defined once the directions of the fields in the relevant terms are known. The present discussion follows that of \cite{Antypas2019IsotopicYtterbium}. We compare the sign of the PV-induced term in (\ref{eq:r0}) with the sign of the term that depends on the $E_{dc}$ field as well as the sign of the PV-mimicking term $b_xe_y\cot\theta$. We consider the harmonics ratio $r_0$ of (\ref{eq:r0})  :

\begin{equation}
\label{eq:r0SignStudies}
r_0=\frac{4E_{dc}}{E_0}+\frac{4\zeta}{\beta E_0}\cot\theta+\frac{4b_xe_y}{B_zE_0}\cot\theta, 
\end{equation}

\noindent where  $b_x$ and $B_z$  are the total fields along x and z, and we have assumed no polarization ellipticity ($\phi=0)$. 

\begin{figure}
\includegraphics[]{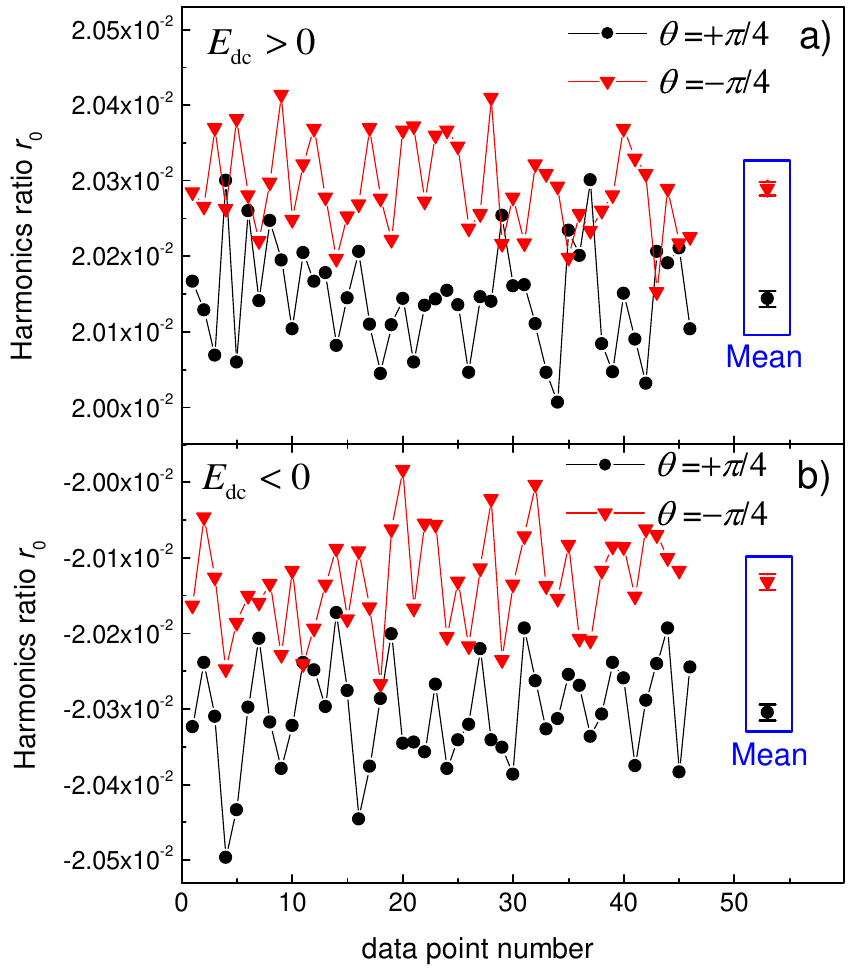}
\caption{(Color online) Harmonics ratio $r_0$ recorded in  $^{172}$Yb over a period of $\approx $25 min. Equal number of data points are shown for either orientation of the leading magnetic field $B_z$, and polarization angles $\theta_{\pm}\approx\pm\pi/4$. In a) the applied $E_{dc}$ field, of approximate magnitude $\vert E_{dc}\vert$=6.3 V/cm, is positive, while in b) it is negative. The ac field applied to the atoms is of amplitude $E_0\approx$1218 V/cm. Observation of the change in $r_0$ with the  $E_{dc}$-reversal establishes a sign definition for $(\zeta/\beta)\cot\theta$. Assuming $\theta$ is set in a way consistent with its sign definition (see coordinate system in fig. \ref{fig:apparatus}), then the observed dependence of $r_0$ on the $\theta$-reversal, is sufficient to determine the $\zeta/\beta$ sign. The angle polarity is checked with measurements presented in fig. \ref{fig:bacyandPV}. The difference in the mean $r_0$ ratios of a) and b) is $-2(8)\cdot10^{-6}$,  corresponding to a difference in the applied $E_{dc}$ magnitudes of $-0.6(5)$ mV/cm,which is indicative of the quality of the $E_{dc}$ reversal in the experiment.} 
\label{fig:r0vsEdc}
\end{figure}

The first step in determining the sign of $\zeta/\beta$ 
is to examine the sign-relationship between the terms $E_{dc}/E_0$ and $(\zeta/\beta)\cot\theta$ in the harmonics ratio $r_0$. Application of a large and positive $E_0$ allows us to adjust the phases of the lock-in amplifiers measuring the $1^{st}$ and $2^{nd}$ harmonics in the excitation rate, to obtain positive outputs with maximal magnitude. (The $E_{dc}$ polarity is checked through measurements made directly on the electric field plates.) We retain these phase values in subsequent PV experiments. We further observe that a reversal of $E_{dc}$ results in a reversal of the $1^{st}$ harmonic sign. With this procedure we establish the convention that $r_0>0$ when $E_{dc}>0$ and $r_0<0$ when $E_{dc}<0$. We then  experimentally check the sign of the extracted term  $(\zeta/\beta)\cot\theta$ in relation to the polarization angle $\theta$. We find that for $\theta>0$, $(\zeta/\beta)\cot\theta<0$ and $(\zeta/\beta)\cot\theta>0$ when $\theta<0$. We show data that illustrate the above observations in fig. \ref{fig:r0vsEdc}. 

\begin{figure}
\includegraphics[]{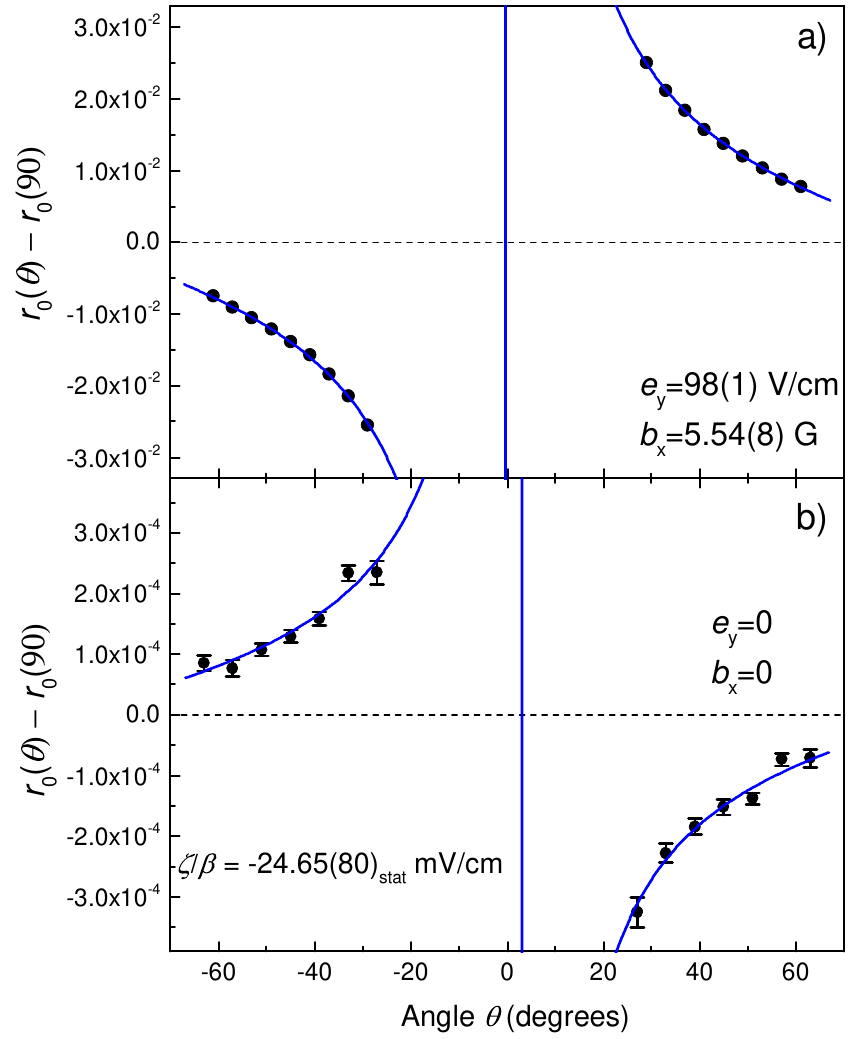}
\caption{(Color online) Harmonics ratio dependence on polarization angle, measured for $^{174}$Yb, with a) application of enhanced field imperfections $b_x$ and $e_y$ (which are both positive as is the 93 G leading field $B_z$), resulting in a dominant contribution to $r_0$ from the term $(b_x/B_z)e_y\cot\theta$, b) without applied imperfections to the atoms, thus primarily observing the PV-related term $(\zeta/\beta)\cot\theta$ . The measured ratio $r_0(\pi/2)$ is subtracted from data taken at  other angles $\theta$, to eliminate background signal which does not depend on $\theta$. Error bars in a) are smaller than data points. The blue line in both plots is a fit to the data of the form $y=y_0+A\cot\ (\theta-\theta_0)$. Observation of the change in the harmonics ratio with $\theta$ in a) offers a confirmation for the sign definition for $\theta$. Furthermore, the $\theta$-dependence of the ratio in b) provides an unambiguous determination of the sign of $\zeta/\beta$. The data of b) also provide a measurement of $\zeta/\beta$, which is consistent with the final result of $-$23.90(11) mV/cm for the $^{174}$Yb isotope.}
\label{fig:bacyandPV}
\end{figure}

The above tests are sufficient to determine the sign of $\zeta/\beta$ provided that the polarization angle is set correctly (see coordinate system in fig. {\ref{fig:apparatus}}). To check this, we extract the contribution of the term $(b_x/B_z)e_y\cot\theta$ in (\ref{eq:r0SignStudies}). With application of enhanced fields $b_x>0$ and $e_y>0$ it is seen that $r_0>0$ for $\theta>0$ and vice versa ($B_z>0$ here). As the polarities of the three relevant fields ($b_x$, $B_z$ and $e_y$) are confirmed before these measurements, we verify that the angle $\theta$ is set correctly, and therefore $\zeta/\beta<0$. Figure \ref{fig:bacyandPV} presents data that support these observations. 

Additional checks were performed to ensure consistency of our sign determination for $\zeta/\beta$. These included a cross-check measurement of the ratio of the \textit{M}1 transition moment and $\beta$, which was determined previously in \cite{Stalnaker2002MeasurementYtterbium}, and independent data analysis of the current PV-data by two of us. These checks are described in \cite{Antypas2019IsotopicYtterbium}.

The sign discrepancy between the present results and the previous Yb PV measurements \cite{Tsigutkin2009ObservationYtterbium, Tsigutkin2010ParitySystematics} was traced to a sign-error in the data analysis performed in that work. The procedure to discover the origin of the discrepancy is discussed in the methods section of \cite{Antypas2019IsotopicYtterbium}.

\subsubsection{\label{sec:level3-Other consistency checks}Other consistency checks}

In addition to the measurements related to the sign of the PV effect, a number of auxiliary  experiments were performed as part of a process to check for unaccounted-for systematics and establish consistency between our model for the harmonics ratio and actual observations under varying apparatus conditions. These experiments are were mentioned in \cite{Antypas2019IsotopicYtterbium}. Here we give a more detailed account of these and provide the respective results in  Table \ref{Table:Aux. Exp. Results}.

The PV isotopic comparison was carried out on the $m=0\rightarrow m'=0$ component of the 408 nm transition, since the systematics in this component are fewer compared to the $m=0\rightarrow  m'=\pm1$ transitions, as shown in Appendix \ref{sec:level1-AppendixA}. However, to verify that the PV-effect is of opposite sign between the $0\rightarrow  0$ and $0\rightarrow \pm1$ transitions, as expected by our model (see Appendix \ref{sec:level1-AppendixA}), data were also taken on the latter components. Indeed, it was observed that the PV effect switches sign between the $0\rightarrow 0$ and $0\rightarrow \pm1$  transitions. To check another expectation, that the PV effect on the $0\rightarrow 0$ transition has a $\cot\theta$ dependence [see eq. (\ref{eq:r0})], the dependence of the harmonics ratio $r_0$ on  the angle $\theta$ was investigated. This expectation was confirmed, as is shown in fig. \ref{fig:bacyandPV}b. The resulting value for $\zeta/\beta$ from these data is consistent with the final measurement for the $^{174}$Yb, shown in the first line of Table \ref{Table:Aux. Exp. Results}.  

Experiments were done using enhanced field imperfection in the interaction region. PV measurements on the $m=0\rightarrow m'=0$ were made with enlarged reversing fields $e^r_y$, $e^r_z$ or non-reversing field $e_z$. The respective results did not reveal unaccounted-for systematics dependent upon  misalignment of the primary electric field, or a stray $e_z$ field, at the $\approx2$\% level. (The effects of the large dc field $E_{dc}$ in the x-direction, as well as a stray field $e_y$, were also studied thoroughly-see sections \ref{sec:level3-Edc and saturation} and  \ref{sec:level3-eybx contribution}.) 

The isotopic comparison data were taken with the 408 nm laser frequency stabilized to the primary peak of the transition lineshape (see fig. \ref{fig:spectrum}b). Systematics related to the transition lineshape   are not expected in the Yb apparatus. Such systematics were present in the Cs PV experiment \cite{Wood1997MeasurementCesium}, owing to the imperfect cancellation of the Stark-$M1$ interference in that work. Here this interference is suppressed to a much greater degree (see Appendix \ref{sec:level1-AppendixA}). To check this expectation, two different experiments were carried out. The first involved  PV-data acquisition on the secondary peak of the $0\rightarrow 0$ lineshape. The result for $\zeta/\beta$ was consistent with that obtained from measurements on the primary peak. The second experiment involved acquiring spectra of the 408 nm transition, such as that shown in fig. \ref{fig:spectrum}, and fitting to the complete lineshape (i.e. to all three lineshape components $0\rightarrow 0,\pm1$, just as it was done in the earlier Yb work \cite{Tsigutkin2009ObservationYtterbium,Tsigutkin2010ParitySystematics}). This method yielded  a $\zeta/\beta$ value  which is consistent with that obtained from the data solely on the $0\rightarrow 0$ transition. The statistical sensitivity in this lineshape-fitting method was lower than that of the measurements with the laser stabilized to the peak of the $0\rightarrow 0$ transition. This is primarily because the impact of laser frequency- noise on measurements done on the side of a peak, is greater than that for data taken at the top of a peak. The effective signal-to-noise-ratio (SNR) in measuring $\zeta/\beta$  with the lineshape-fitting method was $0.06\sqrt{\tau(s)}$, ($\tau$ is the integration time), where as the SNR for PV data taken with the laser locked on the $0\rightarrow 0$ transition was approximately 9 times greater, as discussed in the methods section of \cite{Antypas2019IsotopicYtterbium}.

Further checks with the Yb apparatus were done to confirm a null result for the PV effect in particular cases. A PV effect should not be observed, for instance, when the excitation of atoms is done with circularly polarized light. In such a case, the light ellipticity [see (\ref{eq:RealEopt})] $\phi=\pm\pi/2$, and no PV-related contribution appears in the harmonics ratio $r_0$ (\ref{eq:r0}). A null result was confirmed under such conditions, as shown in Table \ref{Table:Aux. Exp. Results}. Another related experiment made use of the $^{171}$Yb isotope. The ground state $^1$S$_0$ in this nuclear-spin-isotope ($I=1/2$), has a single hyperfine level with   total angular momentum $F=I=1/2$ ( electronic angular momentum $J=0$), where as the excited state $^3$D$_1$ ($J'=1$) has two hyperfine levels with $F'=1/2$ or $3/2$. With application of the typical 93 G magnetic field, the Zeeman sublevels of the excited state are spectrally separated, however the ground state sublevels experience negligible splitting, since $J=0$. Exciting atoms with linearly polarized light to a particular $m_{F'}$ sublevel through the $F=1/2\rightarrow F'=1/2$ transition  (selection rules $\Delta m=0,\pm1$), leads to contributions to the signal from both ground state $m_{F}=\pm1$ levels. As the Stark-PV interference contributions for the two transitions $m_{F}=\pm1/2\rightarrow m_{F'}$ are opposite, no PV observable is expected on the $F=1/2\rightarrow F'=1/2$  transition. A null measurement confirmed this and is presented in Table \ref{Table:Aux. Exp. Results}. 

To obtain additional confidence that the detection of the PV effect if free of spurious apparatus contributions, measurements were done under drastically different conditions. For instance, we took data using a simply constructed set of electric field plates which replaced the elaborate set of electrodes shown in fig. \ref{fig:apparatus}. These measurements are shown in Table \ref{Table:Aux. Exp. Results}.  Another experiment was done with use of a travelling-wave field to excite the 408 nm transition, i.e. without the PBC. The result of the latter measurements is consistent with the final result for the $^{174}$Yb isotope, with the  30\% error being the consequence of poor statistical sensitivity due to the decreased 408 nm optical intensity.

\begin{figure}
\includegraphics[]{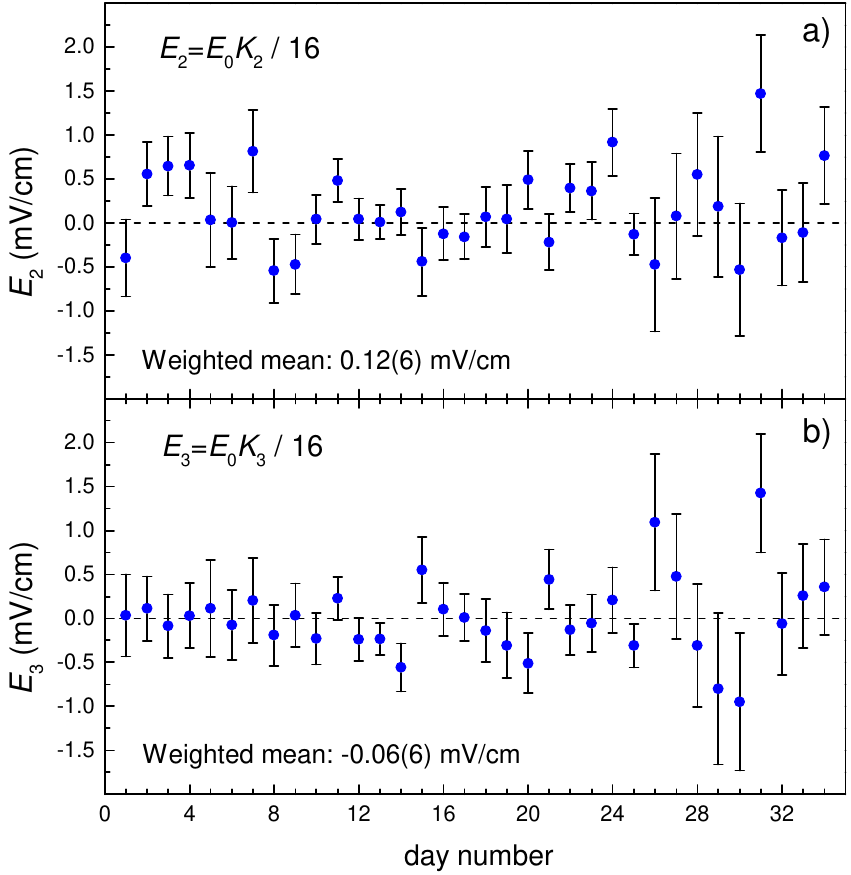}
\caption{(Color online) Effective electric field corresponding to the combination a) $K_2$ and b) $K_3$ (see table \ref{Table:Ktable}). The data come from the isotopic comparison PV-run.}
\label{fig:E2E3}
\end{figure}

Further information about the consistency of the actual isotopic comparison data can be obtained from the analysis of the combinations of Table \ref{Table:Ktable}. The quantity $K_1$ is used to determine the PV effect; $K_4$ is nominally invariant under the $\theta$- and $B$-reversals, and is used to make a precise $E_{dc}$ reversal during data acquisition. The combinations $K_2$ and $K_3$ are related to products of field imperfections (and $\zeta/\beta $) and are expected to be small compared to the measured PV effect. We show in fig. \ref{fig:E2E3} data related to $K_2$ and $K_3$ coming from the actual PV run on the four Yb isotopes. Since data were taken at different electric fields, instead of $K_i$, the effective electric field $E_i=E_0K_i/16$ (i=2,3) is shown in fig. \ref{fig:E2E3}, that can be directly compared to the determined effective PV field $\vert\zeta/\beta\vert\approx$ 23.5 mV/cm. The weighted mean  of $E_3$ is $\approx0.26$\% of $\vert\zeta/\beta\vert$ and consistent with zero  within its $1\sigma$ uncertainty, and the weighted mean of $E_2$ is $\approx0.51$\% of $\vert\zeta/\beta\vert$ and consistent with zero within $2\sigma$.

\begin{table*}[]
\caption{Results of auxiliary experiments}
\begin{ruledtabular}
\begin{tabular}{ c c c c c c }
\textrm{Isotope mass number}&
\textrm{Transition}&
\textrm{Type of experiment}&
\textrm{$\zeta/\beta$ (mV/cm)}\\
\colrule
174 &  $m=0\rightarrow m'=0$ & Actual isotopic comparison data & $-23.89(11)$\\
174 &  $m=0\rightarrow m'=\pm 1$ & $\cdots$ & $23.30(26)$~\footnote{The PV-mimicking terms $e_y^r(e_z/E_0)$ and  $e_z^r(e_y/E_0)$ were not compensated prior to the measurement (see Appendix A and eq. (\ref{eq:AppendixK1fromrm'})).}\\
174 &  $m=0\rightarrow m'=0$ & Measurement of $r_0$  vs. $\theta$~\footnote{See fig. \ref{fig:bacyandPV}. } & $-24.65(80)$\\
174 &  $m=0\rightarrow m'=0$ & Enhanced $e^r_y/E_0=-0.03$ & $-24.30(48)$\\
174 &  $m=0\rightarrow m'=0$ & Enhanced $e^r_y/E_0=0.03$ & $-23.93(40)$\\
174 &  $m=0\rightarrow m'=0$ & Enhanced $e^r_z/E_0=-0.029$ & $-23.98(57)$\\
174 &  $m=0\rightarrow m'=0$ & Enhanced $e^r_z/E_0=0.029$ & $-23.76(57)$\\
174 &  $m=0\rightarrow m'=0$ & Enhanced $e_z/E_0=-0.076$ & $-24.67(57)$\\
174 &  $m=0\rightarrow m'=0$ & Enhanced $e_z/E_0=0.076$ & $-23.83(57)$\\
174 &  $m=0\rightarrow m'=0$ & Measurement on secondary transition peak & $-24.14(44)$\\
174 &  $m=0\rightarrow m'=0, \pm 1$ & Lineshape fitting & $-21(4)$\\
171 &  $F=1/2 \rightarrow F'=1/2$ & $\cdots$ & $-0.59(57)$\\
174 &  $m=0\rightarrow m'=0$ & 408 nm excitation using circularly-polarized light & $-0.2(12)$\\
174 &  $m=0\rightarrow m'=0$ & Measurement with different field plates ~\footnote{Done without the high degree of 408 nm polarization control implemented in the isotopic comparison runs. }  & $-25.2(12)$\\
174 &  $m=0\rightarrow m'=0$ & Measurement without PBC  & $-26(7)$\\
\end{tabular}
\end{ruledtabular}
\label{Table:Aux. Exp. Results}
\end{table*}

\section{\label{sec:level1-Results and analysis}Results and analysis}

In this section we present the results of the PV-isotopic comparison run that took place within a 2.5 month period, from 11/2017 to 01/2018. We compare the observed isotopic variation of the PV effect with the prediction of the SM for this variation. In addition, we present an analysis of these measurements, that is used to constrain electron-nucleon interactions due to the presence of an extra $Z'$ boson. 
\subsection{\label{sec:level2-Results of the PV-measurements}Results of the PV-measurements}
The  results presented here are from data acquired on a chain of four Yb nuclear-spin-zero isotopes, with mass numbers A=170, 172, 174, 176, and abundances 3.1\%, 21.9\%, 31.8\%, 12.9\%, respectively. Measurements were made on the  $m=0\rightarrow m'=0$ component of the 408 nm transition, in 34 days, for a total of 420 hr of integration with $\approx$62 \% duty cycle. A typical routine in the experiment involved loading Yb metal into the oven, studying PV-mimicking systematics, followed by a five day measurement run, of an average 12 hr long daily data-taking time each day. 

The data-acquisition routine was divided in $\approx 30$ min long blocks (PV runs). A PV run consisted of a set of 200 determinations of the harmonics ratio $r_0$, made under all combinations of polarities for the parameters \textit{E}, $\theta$, $B$ and $E_{dc}$. The primary experimental reversal, which is a parity-reversal, was that of the electric field, which was reversed at a rate of 19.9 Hz. The second parity reversal is a $\pm\pi/2$ rotation in $\theta$, which occurred at 0.12 Hz. The primary magnetic field $B$ was reversed at 0.06 Hz and the field $E_{dc}$ at and 0.03 Hz.  The amplitude of the applied ac-field was $E_0\approx 812$ or 1218 V/cm (1218 or 1624 V/cm with $^{170}$Yb). The polarization angle values were $\theta_{\pm}\approx\pm\pi/4$. A total of 884 PV-runs were done, with the number of runs per isotope varying, depending on its respective abundance.  Measurements were alternated among the four spin-zero Yb isotopes, to minimize the impact of potential apparatus drifts.

The measured $\zeta/\beta$ value in each of the four isotopes, is shown in Table \ref{Table:PV-results}. Our  quoted result is the weighted mean of the set of measurements (PV runs) made on the particular isotope. The statistical uncertainty given in Table \ref{Table:PV-results} is the standard error of the respective weighted mean. The systematic uncertainty of 0.06 mV/cm is the same for all isotopes. The  main sources of this uncertainty were discussed in section \ref{sec:level1-Investigation-Systematics}, and their respective contributions are presented in Table \ref{Table:SystErrorsTable}. 

\begin{figure}
\includegraphics[]{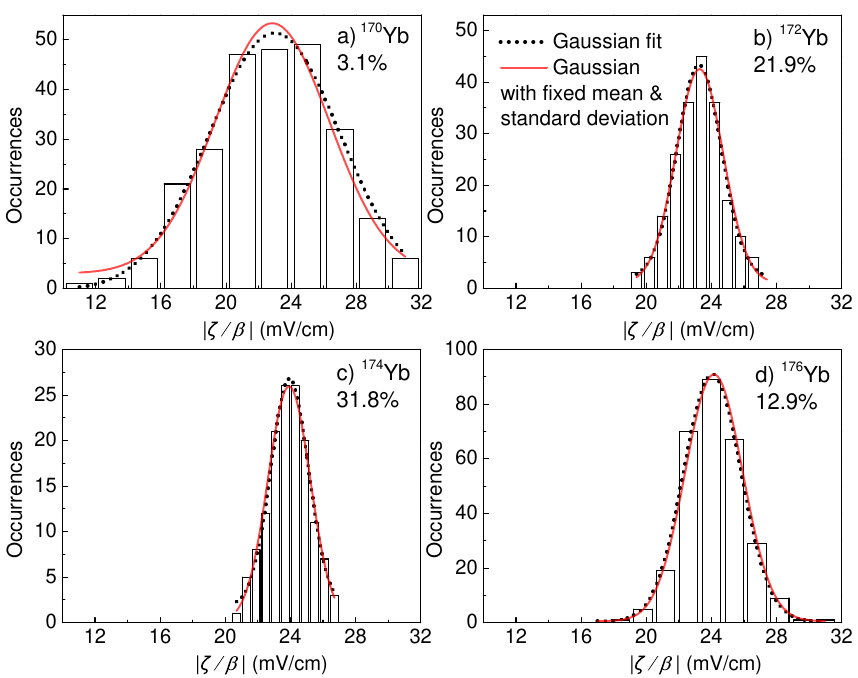} 
\caption{(Color online) Occurrences of $\vert\zeta/\beta\vert$ values observed in 30-min long PV-runs, shown for each on the four Yb isotopes, whose respective abundances are given in the plots.  The distribution of measurements is well-approximated by a Gaussian fit (dotted-black line), indicating that the PV-data are randomly distributed. Also shown in these plots is a solid-red line representing a Gaussian with center given by the mean value from Table \ref{Table:PV-results} and standard deviation which is the standard error of this mean multiplied by $\sqrt{N}$, where $N$ is the number of data points (PV runs) given in Table \ref{Table:PV-results}.}
\label{fig:MeasOccurences}
\end{figure}

 Statistical consistency of the obtained sets of PV data is indicated by the resulting $\chi^2$ value for each isotope, as well as by the probability value associated with the respective set. Consistency of the data is also supported by the frequency count plots of fig. \ref{fig:MeasOccurences}, in which a  random distribution of the measurements is observed.

\begin{table*}[]
\caption{Results of PV measurements}
\begin{ruledtabular}
\begin{tabular}{ c c c c c c }
\textrm{Isotope mass number}&
\textrm{Abundance (\%)}&
\textrm{Number of PV runs}&
\textrm{$\zeta/\beta$ (mV/cm)}&
\textrm{$\chi^2$/d.o.f.}&
\textrm{\textit{p}-value~\footnote{Probability that a repeated experiment would yield a $\chi^2$ value greater than the observed one.}}\\
\colrule
170&3.1 &254& $-22.81(22)_{stat}(0.06)_{syst}$&1.09 & 0.16\\
172&21.9 &199& $-23.24(10)_{stat}(0.06)_{syst}$&0.92 & 0.77\\
174&31.8 &140& $-23.89(11)_{stat}(0.06)_{syst}$&0.99 & 0.53\\
176&12.9 &291& $-24.12(10)_{stat}(0.06)_{syst}$&1.02 & 0.41\\
\end{tabular}
\end{ruledtabular}
\label{Table:PV-results}
\end{table*}

\begin{table}[h]
\caption{Main systematic uncertainties in the PV measurements \cite{Antypas2019IsotopicYtterbium}.}
\begin{ruledtabular}
\begin{tabular}{ l c }
\textrm{Contribution}&
\textrm{Uncertainty(\%)}\\
\colrule
Harmonics ratio calibration & $0.22$\\
Polarization angle & $0.1$\\
High-voltage measurements  & $0.06$\\
Transition saturation correction  &  0.05~\footnote{0.09 for $^{170}$Yb. The error is larger because part of the data for this
low-abundance isotope were taken at a higher electric field.}\\
Field-plate spacing  & $0.04$\\
Stray fields \& field misalignments  & $0.02$\\
Photo-detector response calibration  & $0.02$\\
\end{tabular}
\end{ruledtabular}
\label{Table:SystErrorsTable}
\end{table}

The parameter $\zeta/\beta$ for $^{174}$Yb was reported in \cite{Tsigutkin2009ObservationYtterbium, Tsigutkin2010ParitySystematics} as 39(4)$_{stat}(5)_{syst}$ mV/cm. This magnitude is significantly larger than  that of the present determination for the same isotope, of 23.89$(11)_{stat}(0.06)_{syst}$. The lack of ability to investigate systematics in the apparatus used in \cite{Tsigutkin2009ObservationYtterbium, Tsigutkin2010ParitySystematics} makes it challenging to trace the source of the discrepancy. It is possible that due to the much lower sensitivity of the old apparatus, systematic effects were underestimated.

  The statistical error of the 30 min long PV run varied between  5\% for the highest-abundance isotope ($^{174}$Yb) to 16\% for the lowest-abundance ($^{170}$Yb). The SNR in detection of the PV-effect was $0.55\sqrt{\tau}$ ($\tau$ is the integration time in s) for the highest-abundance isotope. The observed SNR levels in the PV-data acquisition are roughly consistent with shot-noise-limited detection of the 408 nm excitations in the atomic beam. To illustrate this, we compute  the SNR for detection of the Stark-PV interference signal $S_{St-PV}=c_{1}n\zeta \beta E_0$ in the presence of the Stark-induced signal $S_{St}=c_{2}n\beta^2E^2_0$. The parameter $n$ is the atomic beam density, and $c_{1}$, $c_2$ are constants. The noise in detection of $S_{St-PV}$ has three contributions: background ($BG$) noise (independent of $S_{St}$), technical noise $T\cdot S_{St}$ (i.e. proportional to the signal, with $T$ a constant), and shot-noise $S\sqrt{S_{St}}$, where $S$ is a constant. With quadrature addition of these contributions, we obtain for the SNR: 
\begin{multline}
\label{eq:SNRStark-PV}
SNR=\frac{S_{St-PV}}{noise}\\ 
=\frac{c_{1}n\zeta\beta E_0}{\sqrt{BG^2+S^2c_{2}n\beta^2E_0^2+T^2c_2^2n^2\beta^4E_0^4}}.
\end{multline}

\noindent When shot-noise is the dominant noise-source $(S^2c_{1}n\beta^2E_0^2\gg BG^2$, $T^2c_2^2n^2\beta^4E_0^4$),  the SNR is given by:

\begin{equation}
\label{eq:SNRStark-PVShotnoise}
SNR\approx \frac{c_1\zeta}{S\sqrt{c_2}}\sqrt{n}.
\end{equation}
  
\begin{figure}
\includegraphics[]{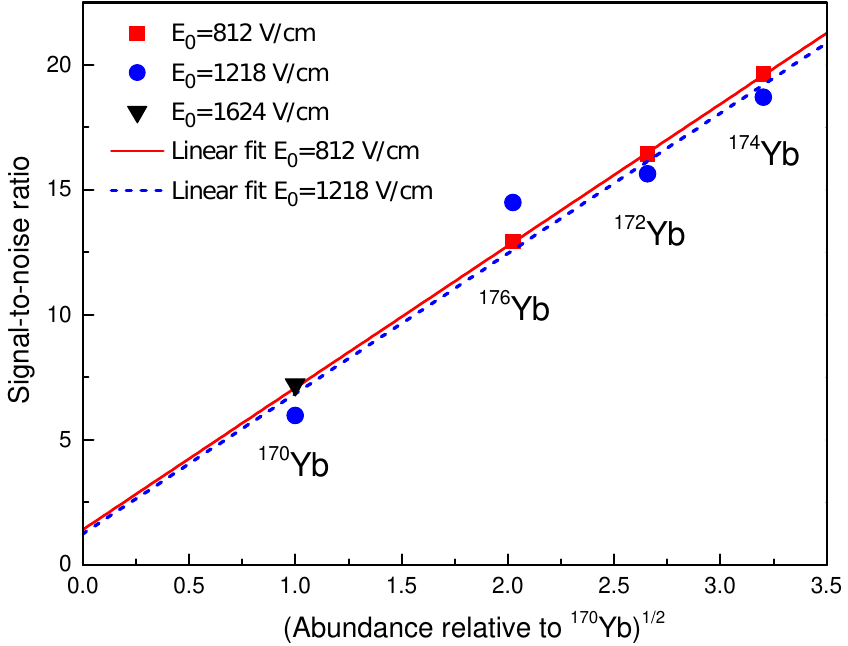}
\caption{(Color online) Obtained SNR in measurements of $\zeta/\beta$ in a 30 min long PV-run, plotted against the square root of the isotopic abundance (or equivalently, the effective atomic beam density). Values are shown for all three different electric fields for which data were taken. Each value in the plot is computed by multiplying the obtained standard error of the weighted mean value $\zeta/\beta$ in the corresponding set of  data by $\sqrt[]{N}$, where $N$ is the number of measurements in the set.}
\label{fig:SNRvsIsotope}
\end{figure}

\noindent We see from (\ref{eq:SNRStark-PVShotnoise}) that the shot-noise-limited SNR does not depend on the electric field $E_0$, and that it scales linearly with $\sqrt[]{n}$. Fig. \ref{fig:SNRvsIsotope} shows the observed SNR of a typical PV run per isotope and per value of $E_0$. The SNR grows approximately as the square root of the isotope abundance, and it has little dependence on the electric field. These observations indicate that the detection of the PV effect approaches the shot-noise limit. Apparatus and measurement method-improvements that resulted in this sensitivity enhancement, relative to that in the $1^{st}$-generation experiment \cite{Tsigutkin2009ObservationYtterbium,Tsigutkin2010ParitySystematics}, are discussed in \cite{Antypas2019IsotopicYtterbium}.

\subsection{\label{sec:level2-Isotopic variation and DM comparison}Isotopic variation of the PV-effect and comparison with SM prediction}

The uncertainty in the present measurements is low-enough to allow for observation of the isotopic variation of the PV effect, and a comparison of this variation with the related prediction of the SM.  The effect predicted by the SM scales as the weak charge of the nucleus $Q_W$, which to lowest order in the SM is given by \cite{Ginges2004ViolationsParticles}:
\begin{equation}
\label{eq:Qweak_lowest_order}
Q_W=-N+Z(1-4\sin^2\theta_W),
\end{equation}
\noindent where $Z,N$ are the number of protons and neutrons in the nucleus and $\theta_W\approx29.2^{\circ}$ is the weak-mixing angle \cite{Tanabashi2018The2018}. A more accurate expression for $Q_W$ \cite{Tanabashi2018The2018} is obtained with inclusion of radiative corrections: 
\begin{equation}
\label{eq:Qweak_with_rad_corr}
Q_W\approx-0.989N+0.071Z.
\end{equation}

\noindent This expression should be accurate at the 0.1\% level. For the mean neutron number $N=103$ of the isotopes measured in this experiment, and $Z=70$, the corresponding weak charge $Q_W=-96.88$, with a proton contribution $Q_p$=4.98. About 95\% of the Yb nucleus weak charge is carried by neutrons. The expected by the SM fractional variation in $Q_W$ per neutron, around N=103 is:
\begin{equation}
\label{eq:FractionalQWchange}
V_{SM}=\frac{1}{Q_{W}}\frac{dQ_W}{dN}\approx1\%.
\end{equation} 

\begin{figure}
\includegraphics[]{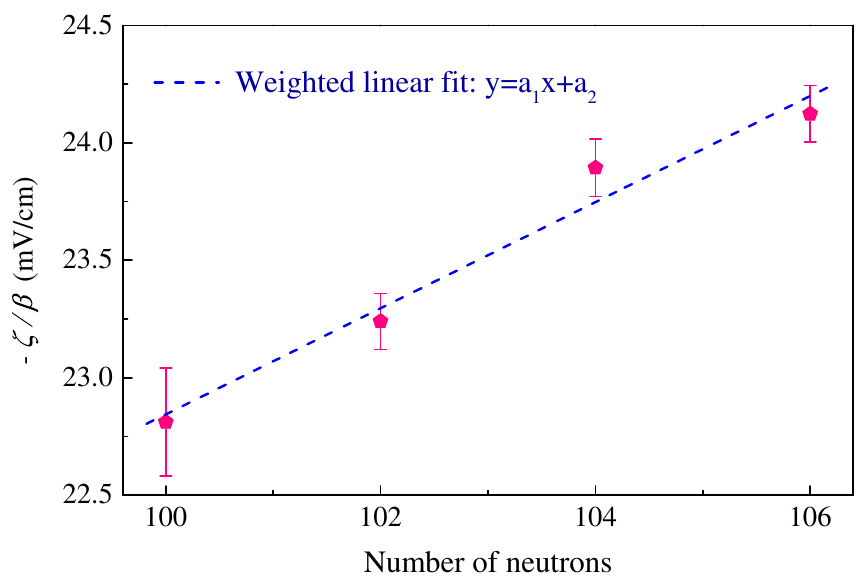}
\caption{(Color online) Isotopic variation of the parameter $-\zeta/\beta$. The error bars shown come from quadrature addition of the statistical and systematic errors and the weight assigned to each point to perform the weighted linear fit to these data is the inverse square of the respective error bar. The parameters of the linear fit are: $a_1=0.226\pm$0.035 mV/cm and $a_2=0.3\pm$3.6 mV/cm. The reduced $\chi^2$ value of the fit is 1.04. Figure adapted from \cite{Antypas2019IsotopicYtterbium}.}
\label{fig:IsotopicComparison}
\end{figure}

A clear variation of the measured PV effect is seen in fig. \ref{fig:IsotopicComparison}, in which the determined $(-\zeta/\beta)$ values for the different isotopes vs. the neutron number are shown. The measured fractional variation in the PV effect per neutron, around N=103, is: 
\begin{equation}
\label{eq:Vexp}
V_{exp}=\frac{\rm slope}{(-\zeta/\beta)_{N=103}}
\end{equation} 

\noindent From the parameters of the fit to the data of fig. \ref{fig:IsotopicComparison}, we obtain  $V_{exp}=0.96(15)\%$. In addition, the y-intercept of the fitted line is consistent with the expected from the SM model contribution due to protons, estimated to be $(Q_p/Q_W)\cdot(- \zeta/ \beta)_{N=103}\approx-1.2$ mV/cm with the value $(-\zeta/\beta)_{N=103}=$23.52 mV/cm obtained from the fit parameters, and $Q_W=-96.88$ for N=103. The small size of the y-intercept is consistent with expectation that the PV effect is mainly due to the neutrons. 

The measured variation of the PV effect $V_{exp}$ agrees well with the SM expectation $V_{SM}$, thus offering a direct confirmation of the $Q_W$ dependence on neutrons.

In determining the variation $V_{exp}$, the effects  of the neutron skin and its variation among the four isotopes measured were neglected. This is reasonable, as the estimated fractional change in the PV effect between the two extreme isotopes $^{170}$Yb and $^{176}$Yb, due to the variation in the neutron skin between these, is only about 0.1\% \cite{Brown2009CalculationsViolation, Viatkina2019DependencePhysics}. This variation is much smaller than the observed change of $\approx$5.7\% between the two extreme isotopes.

\begin{figure}
\includegraphics[width=8.5cm]{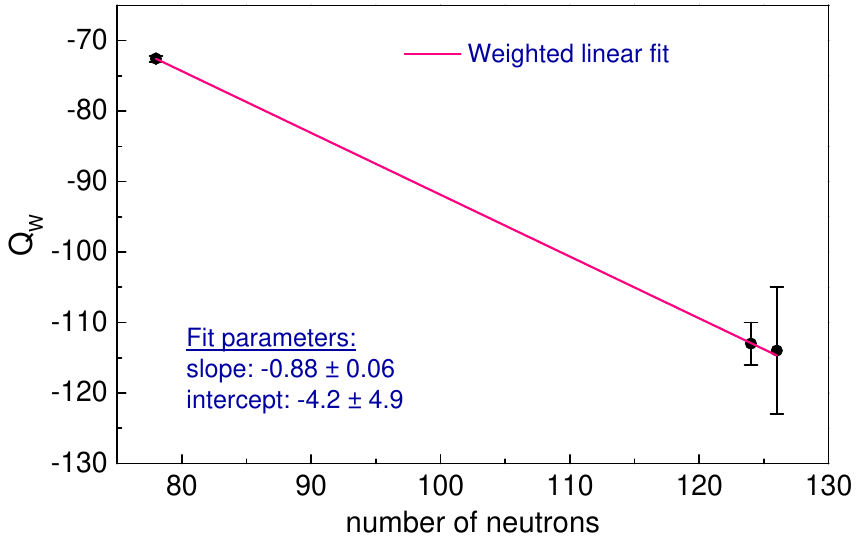}
\caption{(Color online) Nuclear weak charge plotted against number of neutrons. The data points come from the most precise weak charge determinations, made in $^{133}$Cs (\textit{Z}=55, \textit{N}=78), $^{205}$Tl (\textit{Z}=81, \textit{N}=124), and $^{208}$Pb (\textit{Z}=82, \textit{N}=126). The weight assigned to each point in order to perform a weighted linear fit to these data, is the  inverse  square of the  corresponding  error bar shown in the plot. Errors in the fit parameters are the 1$\sigma$ errors.  }
\label{fig:QWvsNZ}
\end{figure}

The most precise determinations of a nuclear weak
charge were made in $^{133}$Cs ($Q_W=-72.58(43)$  \cite{Dzuba2012RevisitingCesium}), $^{205}$Tl ($Q_W=-113(3)$ \cite{Kozlov2001ParityThallium}), and $^{208}$Pb ($Q_W=-114(9)$
through a combination of measurements of the PV effect
with atomic calculations. These determinations combined, have provided a test of the SM regarding the dependence of the weak charge on neutrons and protons. However, taking an agnostic approach, one may question if there is direct evidence of the weak charge being dominated by neutrons. Indeed, we can plot the value of the weak charge inferred from Cs, Tl and Pb experiments, and as a function of the number of neutrons (fig. \ref{fig:QWvsNZ}). The dependence is well fit with a linear function with the slope close to the expected value of $\approx -1$. This fit, however, does not account for correlations in the number of protons and neutrons. To account for such correlations, one can consider instead a weighted fit to the data points of fig. \ref{fig:QWvsNZ}, of the form:

\begin{equation}
\label{eq:DW2Dfit}
 Q_W=A\cdot Z+B\cdot N.
\end{equation} 

\begin{figure}
\includegraphics[width=8.5cm]{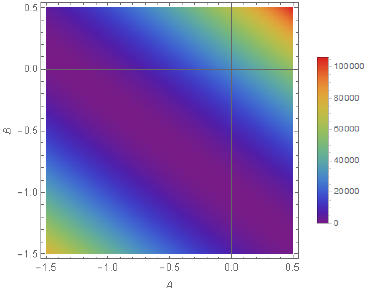} 
\caption{Distribution of the quantity $wss$ of (\ref{eq:WSS}) using the data of fig. \ref{fig:QWvsNZ}. Violet color indicates the region of smaller values for $wss$ and red indicates the region with the largest values within the plotted parameter space. The violet region is a strip extending to infinity.}
\label{fig:QWfitRSS}
\end{figure}

\noindent The poor ability to reliably determine the parameters $A$ and $B$ from such a fit is illustrated in fig. \ref{fig:QWfitRSS}, which shows the distribution of the weighted sum of squares (wss) of the form:
\begin{equation}
\label{eq:WSS}
wss(A,B)=\sum_i\frac{(A\cdot Z_i+B\cdot N_i-Q_{W,i})^2}{\sigma_i^2},
\end{equation} 

\noindent where  $Q_{W,i}$ refer to the data points of fig. \ref{fig:QWvsNZ} and $\sigma_i$ are the respective errors.  One can infer from fig. \ref{fig:QWfitRSS} that a least squares fit  to (\ref{eq:DW2Dfit}) can not provide a reliable estimate for the parameters $A$ and $B$ independently, but rather on the linear combination of $A$ and $B$. Therefore, one can claim that the earlier experiments have not provided a model-independent way of showing that the weak charge is dominated by neutrons and is linear in the number of neutrons, the result we have been able to derive from the isotopic comparison in Yb. To illustrate that the present experiment achieved that, we express the PV-related parameter $\zeta/\beta$ as $\zeta/\beta=k_{at}Q_W$, where $k_{at}$ is a factor which would need to be calculated accurately to extract $Q_W$ from the experiment (see related discussion in section \ref{sec:level1-Introduction}). The quantity $\zeta/\beta$ can be further expressed as:
\begin{equation}
\label{eq:ZetaOverBetakat}
\zeta/\beta=0.2428(A_{exp}\cdot Z+B_{exp}\cdot N)\; \rm{mV/cm},
\end{equation}
\noindent where $k_{at}=0.2428$ mV/cm was computed using the value $Q_W=-96.88$ for N=103 [see eq. (\ref{eq:Qweak_with_rad_corr})]. This value corresponds to  \mbox{$(\zeta/\beta)_{N=103}=-23.52\; \rm{mV/cm}$}, which is extracted from the fit of fig. \ref{fig:IsotopicComparison}. With use of the results of the same fit, we determine the parameters of eq. (\ref{eq:ZetaOverBetakat}): $A_{exp}=-0.01(21)$, $B_{exp}=-0.93(14)$. These values are in agreement with the expected by the SM values [eq. (\ref{eq:Qweak_with_rad_corr})]: $A_{SM}=0.071$ and $B_{SM}=-0.989$. 

\subsection{\label{sec:level2-Constraints on Z' bosons}Constraints on $Z'$ bosons}

The results of the isotopic comparison can be used to place constrain PV couplings between electrons and nucleons that are mediated by an extra vector boson $Z'$. A number of searches for light vector bosons of mass $m_{Z'}>100$ keV,  as well as searches for interactions of SM matter with dark-matter bosons and dark-energy fields have been reported (see, for example, review  \cite{Safronova2018SearchMolecules} and references therein).  Constraints on $Z'$-mediated interactions were placed from torsion-pendulum \cite{Heckel2006NewElectrons,Heckel2008Preferred-frameElectrons} and atomic-magnetometry \cite{Vasilakis2009LimitsComagnetometer} experiments, as well as from atomic calculations \cite{Dzuba2017ProbingMolecules} that employed analyses of results of the Cs PV experiment \cite{Wood1997MeasurementCesium}. These constraints are on combinations of electron-proton and electron-neutron PV interactions. The isotopic-comparison measurements allow for extraction of the proton contribution  to the PV effect. This separation of the electron-proton PV coupling is used to provide individual constraints on an additional electron-proton PV interaction due to  $Z'$ exchange. These new constraints can be combined with existing upper bounds on the sum of electron-proton and electron-neutron couplings, to place a separate limit on electron-neutron interactions. 

The interactions considered here arise in the presence of a $Z'$ boson which does not kinetically mix with the Z boson of the SM. The following $Z'$-mediated interaction is assumed between the electron and nucleons:
\begin{equation}
\label{eq:ZprimeLagrangian}
\mathcal{L}=Z_{\mu}'\sum_{f=e, p, n} \overline{f}\gamma^{\mu}(g_f^V+\gamma_5g_f^A)f,
\end{equation}

\noindent where $Z'_{\mu}$ and $f$ are respectively the boson and fermion amplitudes, and $\gamma^{\mu}$ are Dirac matrices. 

\begin{figure}
\includegraphics[width=8.6cm]{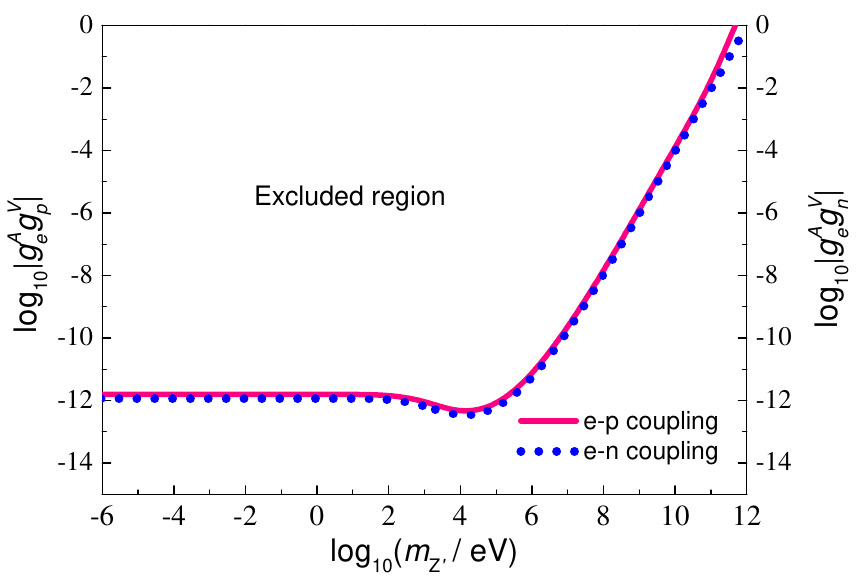}
\caption{(Color online) Constraints on PV electron-nucleon interactions, mediated by an extra $Z'$ boson.  The limit on the axial electron-vector proton interaction $g_e^Ag_p^V$, represented by the pink line, is derived by analysis of the present PV results combined with previous atomic PV calculations \cite{Dzuba2017ProbingMolecules}. The limit on the axial electron-vector neutron interaction  $g_e^Ag_n^V$, represented by the blue dotted line, comes from combination of the obtained bound on $g_e^Ag_p^V$ with a previous result on an effective electron-nucleon coupling. The latter coupling was derived in \cite{Dzuba2017ProbingMolecules} through analysis of the results of \cite{Wood1997MeasurementCesium}.  Both bounds are shown at the 67\% confidence level. The low-mass and high-mass asymptotic limits for these bounds are given in Table \ref{Table:ZprimeExclusions}. Figure adapted from \cite{Antypas2019IsotopicYtterbium}.}
\label{fig:ZprimeExclusions}
\end{figure}

The present isotopic comparison data provide an estimate for the proton contribution to the PV parameter $\zeta/\beta$. This estimate is used in combination with the atomic calculations of \cite{Dzuba2017ProbingMolecules} to place the upper bound on the axial electron-vector proton coupling $g^A_eg^V_p$. The atomic PV calculations reported in \cite{Dzuba2017ProbingMolecules} assume a finite mass for the $Z'$ boson. Therefore the corresponding couplings and bounds of these are defined for any mass $m_{Z'}$, and not only in the limit of a mass which is large on the atomic scale. The bound obtained on $g^A_eg^V_p$ is combined with a previous bound on an effective electron-nucleon coupling $g^A_eg^V_N$ (see analysis in \cite{Dzuba2017ProbingMolecules}) to constrain the axial electron-vector neutron coupling $g^A_eg^V_n$. A detailed account of the analysis to derive bounds on $g^A_eg^V_p$ and $g^A_eg^V_n$ is given in \cite{Antypas2019IsotopicYtterbium}, and here we  only provide its main results. We show in fig. \ref{fig:ZprimeExclusions} the constraints derived on the $Z'$-mediated electron-proton and electron-neutron couplings. In Table \ref{Table:ZprimeExclusions} we present all the asymptotic values for the couplings $g^A_eg^V_p$ and $g^A_eg^V_n$ in the limits of low mass and high mass for $Z'$.

\begin{table*}[]
\caption{ Upper bounds on electron-proton and electron-neutron interactions mediated by a vector boson $Z'$ of mass $m_{Z'}$. These limits are derived through analysis of the results of different experiments or combinations of these. The large-mass limits $ g_e^Ag_p^V/m_{Z'}^2$ and $ g_e^Ag_n^V/m_{Z'}^2$ are valid for  $m_{Z'}>$ 1 GeV and the low-mass limits $ g_e^Ag_p^V$ and $ g_e^Ag_n^V$ for $m_{Z'}<$ 100 eV (see table I in \cite{Dzuba2017ProbingMolecules}). Table reproduced from supplementary material of \cite{Antypas2019IsotopicYtterbium}. }
\begin{ruledtabular}
\begin{tabular}{ l c c c c c }
\textrm{}&
\textrm{$ g_e^Ag_p^V/m_{Z´}^2$ (GeV)$^{-2}$}&
\textrm{$ g_e^Ag_n^V/m_{Z´}^2$(GeV)$^{-2}$}&
\textrm{$ g_e^Ag_p^V$}&
\textrm{$ g_e^Ag_n^V$}\\
\textrm{}&
\textrm{ (\textit{large-mass limit})}&
\textrm{(\textit{large-mass limit})}&
\textrm{(\textit{low-mass limit})}&
\textrm{(\textit{low-mass limit})}\\

\textrm{Experiment}&
\textrm{}&
\textrm{}&
\textrm{}&
\textrm{}\\
\colrule 
Yb PV & $(3.7\pm 9)\cdot 10^{-7}$ & $\cdots$ &$(4.5\pm 11)\cdot 10^{-13}$  & $\cdots$ \\[0.2cm]
Yb \& Cs PV & $\cdots$ & $(-2.9\pm 6.4)\cdot 10^{-7}$ & $\cdots$ & $(-3.5\pm 7.9)\cdot 10^{-13}$ \\[0.2cm]
Q-weak & $(-4.5\pm 18.6)\cdot 10^{-9}$& $\cdots$ & $\cdots$ &  $\cdots$\\[0.2cm]
Q-weak \& Cs PV & $\cdots$&$(-3.1\pm 2.6)\cdot 10^{-8}$  & $\cdots$ & $\cdots$ \\[0.2cm]

\end{tabular}
\end{ruledtabular}
\label{Table:ZprimeExclusions}
\end{table*}

\section{\label{sec:level1-Conclusions and outlook}Conclusions and outlook}

We discussed in detail the experimental principle used to make improved measurements of the PV effect in a chain of four Yb isotopes. We described the $2^{nd}$-generation atomic beam apparatus, which offers enhanced sensitivity in the detection of the PV effect, thus enabling better characterization of systematic effects in these measurements. We gave a detailed account of the studies of these systematic effects, in relation to the isotopic-comparison experiment. 

 The results of the PV measurements presented here offer the first direct observation of isotopic dependence in atomic PV. The measured variation in the PV effect, of $0.96(15)\%$ per neutron, is in agreement with the expectation based on the electroweak theory, of $\approx1\%$ per neutron. Our result is consistent with the notion of the magnitude of the neutron weak charge being close to unity (Eq. (\ref{eq:Qweak_lowest_order})) and the weak charge of the nucleus to be additive over the neutrons.  
 
 The isotopic-comparison method  allowed the extraction of the proton contribution to the PV effect. This contribution has enabled analysis that provided constraints on axial electron-vector proton interactions, mediated by a light boson $Z'$. These new constraints were combined with existing constraints on the sum of electron-proton and electron-neutron couplings, to  provide separate constraints on the latter. 
 
 The attained single-isotope uncertainty is $\approx0.5\%$ for three of the Yb isotopes measured. The present sensitivity level is a  benchmark for the newly-built apparatus. Many avenues to enhance sensitivity have been identified and are currently being explored. These include an upgrade of the PBC cavity optics for a greater circulating power level, an optimization in the atomic beam flux, potentially involving laser cooling of the transverse velocity distribution of atoms. 
 
 A tenfold sensitivity enhancement should allow a measurement of the variation of neutron distributions in the Yb nucleus with use of the isotopic comparison method \cite{Dzuba1986EnhancementAtoms, Viatkina2019DependencePhysics}.
 A tenfold sensitivity increase is also expected to be sufficient for an anapole moment measurement. The nuclear-spin-dependent PV amplitude, which is active for isotopes with nuclear spin ($^{171}$Yb, $I=1/2$, $^{173}$Yb, $I=5/2$), contributes by $\approx$ 0.1\% to the overall PV effect \cite{V.V.Flambaum1984PossibilityExperiments, Porsev2000ManifestationYtterbium,Singh.A.D.1999ParityMoment,Dzuba2011CalculationYtterbium} but this contribution depends on the particular hyperfine level. PV measurements on different hyperfine levels in the same fermionic isotope are therefore required to probe an anapole. For instance, an anapole extraction can be done by measuring the difference in the PV amplitudes between the $F=1/2\rightarrow F'=1/2$ and $F=1/2\rightarrow F'=3/2$ transitions in $^{171}$Yb, or between the $F=5/2\rightarrow F'=3/2$ and $F=5/2\rightarrow F'=7/2$ transitions in $^{173}$Yb. Optical pumping to an extreme magnetic sublevel of the Yb ground state, will improve statistical sensitivity and simplify the analysis of systematics. This pumping is necessary, in order to obtain a PV observable on the $F=1/2\rightarrow F'=1/2$ component of  $^{171}$Yb (see discussion in section \ref{sec:level3-Other consistency checks}).

An increase in the experimental sensitivity must be accompanied by improved understanding and control of systematic effects. With consideration to improved isotopic comparison measurements on a chain of nuclear-spin-zero isotopes, systematic effects should not pose substantial difficulties. This is because the energy level structure is identical for the different isotopes, and since the influence of such spurious effects on measurements made on the $\Delta m=0$ transition is only moderate. The various calibrations applied to the data, as well as the associated uncertainties, are also largely independent of isotope measured. 

Greater attention to systematics is required in the studies of spin-dependent PV. It is possible that some effects could contribute differently among the different hyperfine transitions, and affect the results of hyperfine comparison. A substantial amount of related studies was done in the Cs experiment \cite{Wood1997MeasurementCesium}, which (similarly to the present work) employed the Stark-PV interference method and was done with an atomic beam, with the use of a standing-wave field to excite atoms. Systematic contributions influencing the hyperfine comparison in that work came from the presence of a \textit{M}1 transition amplitude, which, although suppressed due to the use of a standing-wave, was allowed by the geometry of applied fields. In the Yb experiment, in addition to the suppression provided by the PBC, the experimental geometry is such that the Stark and \textit{M}1 amplitudes are out of phase for the $\Delta m=0$ transition that we employ, and therefore they do not interfere. In addition to \textit{M}1-related systematics, the effects of an electric-quadrupole (\textit{E}2) transition between the $^1$S$_0$ and $^3$D$_1$ states need to be considered. The \textit{E}2 transition is weakly allowed in the nonzero-spin isotopes due to hyperfine interaction-induced mixing between the $^3$D$_1$ and $^3$D$_2$ states. A detailed evaluation of the \textit{E}2 amplitudes in the  $^1$S$_0\rightarrow ^3$D$_1$ transition was reported in  \cite{Kozlov2019Hyperfine-induced1S0-3D1}. Fortunately, the same mechanisms employed to suppress the Stark and \textit{M}1 effects in PV measurements (experimental field geometry, excitation with counter-propagating light beams, selection of $\Delta m=0$ transitions), are expected to provide adequate suppression of Stark-\textit{E}2 signal contributions in the nonzero-spin isotopes.  Modeling of systematics in the these isotopes, just as it was done for the studies presented here,  shows that parasitic contributions to the true PV signal should be similar to those in the spin-zero isotopes. While the analysis indicates that it should be possible to control systematics in the measurements of the nuclear-spin-dependent PV, we expect that during the course of the Yb PV program, studies of systematics will require most of our attention.

\section*{Acknowledgements}
We are gratetful to M. Safronova, M. Kozlov, S. Porsev, M. Zolotorev, A. Viatkina, Y. Stadnik, L. Bougas and N. Leefer for
fruitful discussions. VF thanks  Gutenberg Fellowship and Australian Research Council. AF is supported by the Carl Zeiss Graduate Fellowship.

\bibliography{Mendeley}

\appendix

\section{\label{sec:level1-AppendixA}Transition rates and  ratios $r_{m'}$ for the $m=0\rightarrow m'=0,\pm 1$ transitions}

The amplitudes for the Stark-,  PV-induced and magnetic-dipole (\textit{M}1) transitions between the $^1$S$_0, m=0\rightarrow ^3$D$_1,m'=0,\pm1$ states are given by \cite{Tsigutkin2010ParitySystematics}:
\begin{equation}
\label{eq:AppendixAstarkm}
A^{Stark}_{m'}=i\beta(-1)^{m'}(\vec{E}\times\vec{\mathcal{E}})_{-m'},
\end{equation}
\begin{equation}
\label{eq:AppendixAPVm}
A^{PV}_{m'}=i\zeta(-1)^{m'}\vec{\mathcal{E}}_{-m'},
\end{equation}
\begin{equation}
\label{eq:AppendixAM1m}
A^{M1}_{m'}=M1(-1)^{m'}(\hat{k}\times\vec{\mathcal{E}})_{-m'},
\end{equation}

\noindent where $\hat{k}$ is the unit vector related to the optical field with electric field amplitude $\vec{\mathcal{E}}$, and \textit{M}1 is the magnetic-dipole transition moment, measured in \cite{Stalnaker2002MeasurementYtterbium}: $\vert M1\vert=1.33(21)\times 10^{-4} \mu_B$, where $\mu_B$ is the Bohr magneton. The geometry of applied fields in the present experiment, however, is such that $A^{M1}$ is nominally (i.e. in the absence of experimental imperfections) out of phase with respect to $A^{Stark}$, so that the two amplitudes do not interfere. Later in this Appendix we present the leading PV-mimicking systematic due to residual Stark-$M1$ interference. 

A second level of  suppression  of the effects of the $M1$ amplitude is due to the standing-wave nature of the optical field in the PBC \cite{Bouchiat1981Can}. As discussed in \cite{Tsigutkin2010ParitySystematics}, the \textit{M}1 amplitude in the presence of a standing-wave with counter-propagating field amplitudes $\vec{\mathcal{E}}_+$ and $\vec{\mathcal{E}}_-$, is given by:

\begin{equation}
\label{eq:AppendixAM1mStandingWave}
A^{M1,PBC}_{m'}=M1(-1)^{m'}(\kappa \hat{k}\times\vec{\mathcal{E}})_{-m'},
\end{equation}

\noindent where $\vec{\mathcal{E}}=\vec{\mathcal{E}}_++\vec{\mathcal{E}}_-$ and $\kappa=(\mathcal{E}_+- \mathcal{E}_-)/ \mathcal{E}$. The amplitude of (\ref{eq:AppendixAM1mStandingWave}) is suppressed by a factor 1/$\kappa$, relative to that induced by a traveling-wave field. The suppression is $\approx$ 300 in the present experiment.   

We first consider the ideal case, in which there are no stray-fields, field misalignments, or ellipticity in the optical field polarization and $\kappa=0$, and derive expressions for the three frequency components $R_{m'}^{[0]}$, $R_{m'}^{[1]}$, and $R_{m'}^{[2]}$   present in the excitation rate $R_{m'}$. In this case, the electric, magnetic, and optical fields are as follows: 
\begin{equation}
\label{eq:E_staticAppendixIdeal}
\vec{E} =(E_{dc} +E_0\cos\omega t )\hat{x},
\end{equation}
\begin{equation}
\label{eq:BAppendixIdeal}
\vec{B} =B_z\hat{z},
\end{equation}
\begin{equation}
\label{eq:EopticalAppendixIdeal}
\vec{\mathcal{E}}=\mathcal{E}(\sin\theta \hat{y}+\cos\theta\hat{z}).
\end{equation}
\noindent The transition rate $R_{m'}$ is given by:
\begin{align}
\label{eq:Rm'Appendix}
R_{m'}\propto\vert A_{m'}^{Stark}+A_{m'}^{PV}\vert ^2 \\
=R_{m'}^{[0]}+R_{m'}^{[1]}\cos\omega t +R_{m'}^{[2]}\cos2\omega t.
\end{align}
\noindent Evaluating the amplitudes of (\ref{eq:AppendixAstarkm}, \ref{eq:AppendixAPVm} and \ref{eq:AppendixAM1mStandingWave}) in (\ref{eq:Rm'Appendix}) yields the following harmonic amplitudes for the $0\rightarrow 0$ transition rate:
\begin{multline}
\label{eq:AppendixR0_0}
R_0^{[0]}\approx2\mathcal{E}^2\beta^2E_0^2\sin^2\theta+4\mathcal{E}^2\beta^2E_{dc}^2\sin^2\theta \\+8\mathcal{E}^2\beta E_{dc}\zeta\cos\theta\sin\theta,
\end{multline}
\begin{equation}
\label{eq:AppendixR1_0}
R_0^{[1]}=8\mathcal{E}^2\beta E_0\zeta\cos\theta \sin\theta+8\mathcal{E}^2\beta^2 E_0E_{dc}\sin^2\theta,
\end{equation}
\begin{equation}
\label{eq:AppendixR2_0}
R_0^{[2]}\approx2\mathcal{E}^2\beta^2 E_0^2\sin^2\theta.
\end{equation}
\noindent The amplitudes for the $0\rightarrow \pm1$ transitions are: 
\begin{multline}
\label{eq:AppendixR0_m'}
R_{\pm1}^{[0]}\approx\mathcal{E}^2\beta^2E_0^2\cos^2\theta+2\mathcal{E}^2\beta^2E_{dc}^2\cos^2\theta \\-4\mathcal{E}^2\beta E_{dc}\zeta\cos\theta\sin\theta,
\end{multline}
\begin{equation}
\label{eq:AppendixR1_m'}
R_{\pm1}^{[1]}=-4\mathcal{E}^2\beta E_0\zeta\cos\theta \sin\theta+4\mathcal{E}^2\beta^2 E_0E_{dc}\cos^2\theta,
\end{equation}
\begin{equation}
\label{eq:AppendixR2_m'}
R_{\pm1}^{[2]}=\mathcal{E}^2\beta^2 E_0^2\cos^2\theta.
\end{equation}

\noindent The terms proportional to $\zeta^2$ were dropped in (\ref{eq:AppendixR0_0}, \ref{eq:AppendixR2_0}, \ref{eq:AppendixR0_m'}). The apparatus measures the ratio of the 1$^{st}$- to the 2$^{nd}$ harmonic in the transition rate. When exciting the $0 \rightarrow 0$ transition, this ratio is given by: 
\begin{equation}
\label{eq:Appendixr0ideal}
r_0\equiv \frac{R_0^{[1]}}{R_0^{[2]}}=\frac{4E_{dc}}{E_0}+\frac{4\zeta}{\beta E_0}\cot\theta.  
\end{equation}

\noindent For the $0 \rightarrow\pm 1$ transition, the corresponding ratio is: 
\begin{equation}
\label{eq:Appendixrm'ideal}
r_{\pm1}\equiv \frac{R_{\pm1}^{[1]}}{R_{\pm1}^{[2]}}=\frac{4E_{dc}}{E_0}-\frac{4\zeta}{\beta E_0}\tan\theta. 
\end{equation}
  
We now derive expressions for $r_0$ and $r_{\pm1}$ in the presence of apparatus imperfections. In this case the fields $\vec{E} , \vec{B}$ and $\vec{\mathcal{E}}$ are expressed as: 
\begin{equation}
\label{eq:AppendixRealE}
\vec{E}=(E_{dc}+E_0\cos \omega t)\hat{x}+(e_{y}+e_y^r\cos \omega t)\hat{y}+(e_z+e_{z}^r\cos \omega t)\hat{z},
\end{equation}
\begin{equation}
\label{eq:AppendixRealB}
\vec{B}=(b_x+f_B b_x^r)\hat{x}+(b_y+f_B b_y^r)\hat{y}+(b_z+f_B B_z)\hat{z},
\end{equation}
\begin{equation}
\label{eq:ApendixRealEopt}
\vec{\mathcal{E}}=\mathcal{E}(\sin\theta \hat{y}+\cos\theta e^{i\phi} \hat{z}).
\end{equation}
\noindent We include in this analysis the contribution of the M1 amplitude (eq. \ref{eq:AppendixAM1mStandingWave}). The various terms in (\ref{eq:AppendixRealE}, \ref{eq:AppendixRealB}, \ref{eq:ApendixRealEopt}) were introduced in section \ref{sec:level1-Experimental Method}. The presence of the small $b'_x=b_x+f_B b_x^r$ and $b'_y=b_y+f_B b_y^r$ components, in addition to the leading field $B'_z=b_z+f_B B_z$, is responsible for Zeeman mixing of adjacent $m'$ sublevels of the $^3$D$_1$ state, which needs to be considered in deriving the expressions for the transition rate $R_{m'}$ and the harmonics ratios $r_{m'}$. One approach is to compute this mixing and modify the amplitudes of (\ref{eq:AppendixAstarkm}, \ref{eq:AppendixAPVm} and \ref{eq:AppendixAM1m}) accordingly. Alternatively, the fields $\vec{E} , \vec{B}$ and $\vec{\mathcal{E}}$ can be rotated by application of an operator $\mathbf{\mathcal{D}}=\textit{D}(-a_y,\hat{y})\textit{D}(a_x,\hat{x})$, such that $\mathbf{\mathcal{D}}\vec{B}\propto\hat{z}$ \cite{Tsigutkin2010ParitySystematics}. The rotation angles are $a_{x(y)}=b'_{y(x)}/B'_z$. The rotated fields $\mathcal{D}\vec{E}$ ,  $\mathcal{D}\vec{B}$, $\mathcal{D}\hat{k}$ and $\mathcal{D}\vec{\mathcal{E}}$ are used to evaluate the transition amplitudes of eqns (\ref{eq:AppendixAstarkm}, \ref{eq:AppendixAPVm} and \ref{eq:AppendixAM1m}). A great number of terms appear then in the expression for the rates $R_{m'}$. The corresponding harmonics ratios $r_{m'}$ are expanded in terms of the small field imperfections, the parameter $\kappa$ and $\zeta$. The ratios  $r_{0}$ and  $r_{\pm1}$, retaining terms up to 2$^{nd}$ order in the expansion, are as follows:

\begin{multline}
\label{eq:Appendixr0}
r_0(\theta,f_B)=\frac{4E_{dc}}{E_0}-\frac{4(b_x+f_Bb_x^r)e_z}{f_BB_zE_0}+\\ 
\left[\frac{4\zeta}{\beta E_0}+\frac{4(b_x+f_Bb_x^r)e_y}{f_BB_zE_0}\right]\cot\theta \cos\phi \\ +\frac{4(b_y+f_Bb_y^r)\zeta}{f_BB_zE_0}(1+\cot^2\theta \cos 2\phi),
\end{multline}

\begin{multline}
\label{eq:Appendixrm'}
r_{\pm 1}(\theta,f_B)=\frac{4E_{dc}}{E_0}+\frac{4e^r_ye_y}{E_0^2}\pm\frac{4e^r_yM1\kappa}{\beta E_0^2}+\frac{4(b_y+f_Bb^r_y)\zeta}{f_BB_z\beta E_0}+\\
\left[-\frac{4\zeta}{\beta E_0}-\frac{4(b_x+f_Bb_x^r)e_y}{f_BB_zE_0}-\frac{4e_y^re_z}{E_0^2}-\frac{4e_z^re_y}{E_0^2}\mp \frac{4e^r_zM1\kappa}{\beta E_0^2}
\right]\\
\times\tan\theta\cos\phi  \pm\frac{4e_z}{E_0}\tan\theta\sin\phi + \\
\left[ \frac{4(b_x+f_Bb^r_x)e_z}{f_BB_zE_0}+\frac{4e_z^re_z}{E_0^2}+\frac{4b_y\zeta}{f_BB_zE_0}\right]\tan^2 \theta\cos2\phi
\end{multline}
\\
\noindent To compare the sensitivity of $\zeta/\beta$  measurements, made in different transition components, to false-PV effects, we form the combination $K_1$ [see eq. (\ref{eq:Ki_Matrix_equation}) and Table \ref{Table:Ktable}] with use of $r_0$ and the sum $(1/2)(r_{-1}+r_{+1})$:

\begin{equation}
\label{eq:AppendixK1fromr0}
K_1^{0}=\left[\frac{8\zeta}{\beta E_0}+\frac{8b_x^re_y}{B_zE_0}\right](\cot\theta_+\cos\phi_+-\cot\theta_-\cos\phi_-),
\end{equation}

\begin{multline}
\label{eq:AppendixK1fromrm'}
K_1^{\pm1}=\left[-\frac{8\zeta}{\beta E_0}-\frac{8b_x^re_y}{B_zE_0}-\frac{8e^r_ye_z}{E_0^2}-\frac{8e^r_ze_y}{E_0^2}\right]\times \\
(\tan\theta_+\cos\phi_+-\tan\theta_-\cos\phi_-).
\end{multline}

\noindent We see that there are more false-PV terms in  $K_1^{\pm1}$, compared to $K_1^{0}$. A misalignment such that $e^r_y/E_0=0.005$, for instance, coupling to a stray $e_z$=50 mV/cm, gives rise to a PV-mimicking signal of 0.25 mV/cm, which is $\approx$ 1\% of the measured PV effect. This is the primary reason why the isotopic comparison data were taken at the $0\rightarrow 0$ transition component. 

We now evaluate the impact of the $M1$-related PV-mimicking contributions, to separately illustrate the effectiveness of the two methods used to suppress the effects of the magnetic dipole transition, namely the choice of experimental field geometry  and the excitation of the 408 nm transition with a standing-wave field. We focus on the $m=0\rightarrow m'=0$ component of the $^1$S$_0\rightarrow^3$D$_1$ transition, however, analysis on the $m=0\rightarrow m'=\pm1$ components yields similar conclusions. We start by evaluating the suppression provided by the experimental field geometry. For this, we compute the harmonics ratio $r_0(
\theta, f_B)$ and the associated combination $K_1^0$, assuming that a traveling-wave field excites atoms, i.e. the parameter $\kappa$ of (\ref{eq:AppendixAM1mStandingWave}) is not negligible. If the  ratio $r_0(\theta, f_B)$ is expanded in terms of the various field imperfections, a third order term in the small parameters that contains the $M1$ amplitude appears in $r_0(\theta, f_B)$, which mimics the PV-term. We omit the number of contributions from Stark-induced systematics, and focus on the competition between the M1- and PV-related signals. The relevant part of the combination $K_1^0$ is: 
\begin{multline}
\label{eq:AppendixK1fromr0M1effects}
K_1^{0}=\Big(\frac{8\zeta}{\beta E_0}\Big)(\cot\theta_+\cos\phi_+-\cot\theta_-\cos\phi_-)\\
-\Big(\frac{8b^r_xe_y^r\kappa M1}{B_z\beta E_0^2}\Big)(\cot\theta_+\sin\phi_+-\cot\theta_-\sin\phi_-).
\end{multline}

\noindent Let us form the ratio $r_{M1-PV}$ of the $M1$- and  PV-related contributions to $K_1^0$:

\begin{multline}
\label{eq:rM1PV}
r_{M1-PV}=-\Big(\frac{M1/\beta}{\zeta/\beta }\Big)\Big(\frac{b^r_xe^r_y}{B_zE_0}\Big)\kappa \\
\times \frac{\cot\theta_+\sin\phi_+-\cot\theta_-\sin\phi_-}{\cot\theta_+\cos\phi_+-\cot\theta_-\cos\phi_-}.
\end{multline}

\noindent The ratio $M1/\beta\approx -22.3$ V/cm \cite{Stalnaker2002MeasurementYtterbium}, $\zeta/\beta\approx -$23.9 mV/cm, and $\cot\theta_{\pm}\approx\pm1$. Assuming reasonable values for imperfections: $b^r_x/B_z=10^{-4}$, $e_y^r/E_0=0.005$, $\phi_{\pm}=0.05$ rad, we obtain $r_{M1-PV}=-2.3\cdot10^{-5}\kappa$. Therefore, the present choice of experimental field geometry is sufficient to provide a practically complete suppression of the contribution of the M1-related systematic, even if the experiment were to be carried out with a travelling-wave field to excite atoms ($\kappa=1$). The standing-wave field in the PBC provides further suppression ($\kappa\approx1/300$), resulting in a residual fractional contribution of the  PV-mimicking signal due to the $M1$ amplitude of $\approx7.8\cdot10^{-8}$.

In section \ref{sec:level3-polarization parameter p} we discussed the measurement of the polarization ellipticity-related angle $\phi$. This angle is determined by combining measurements of the difference $r_{+1}-r_{-1}$, made for opposite polarities of an enhanced $e_z$ field $(\pm e_z)$. Use of (\ref{eq:Appendixrm'}) in this case yields:
\begin{multline}
\label{eq:AppendixEllipticityfromrm}
(r_{+1}-r_{-1})_{+e_z}-(r_{+1}-r_{-1})_{-e_z}=\\
\frac{16e_z}{E_0}\tan\theta\sin\phi
\end{multline}

\noindent This expression was used in the analysis presented in section \ref{sec:level3-polarization parameter p}. 

\section{\label{sec:level1-AppendixB}Measuring $\theta_{\pm}$ using the 408 nm profile}

Here we describe the method to measure the polarization angles $\theta_{\pm}$ using recorded profiles of the 408 nm resonance. These measurements are correlated with the concurrent readings of a polarimeter monitoring the light transmitted through the PBC, whose subsequent readings during a PV run are used to provide  continuous tracking of the $\theta_{\pm}$ angles.

 The polarization angles input to the PBC are set to approximately $\pm\pi/4$ (i.e. to the nominal values for which PV data are acquired) and are determined  through analysis of the relative peak heights for the three transition components $m=0\rightarrow m'=0,\pm1$. Let $R_{0}^{[2]'}, R_{\pm1}^{[2]'}$ be the $2^{nd}$ harmonic amplitudes of the  $0\rightarrow 0$ and $0\rightarrow\pm1$ transitions, given by:
\begin{equation}
R_{0}^{[2]'}=R_{0}^{[2]}+h(R_{-1}^{[2]}+R_{+1}^{[2]}),
\end{equation}
\begin{equation}
R_{\pm1}^{[2]'}=R_{\pm1}^{[2]}+hR_{0}^{[2]},
\end{equation}
\noindent where $R_{m'}^{[2]}$ is the amplitude of the $m'$ transition component in the absence of peak overlap, and $h=0.00042(4)$ [introduced in (\ref{eq:Rpeak_overlap})] is a parameter quantifying the slight overlap of adjacent peaks in the spectrum. The amplitudes $R_{m'}^{[2]}$  include a small correction for the slight saturation of the corresponding transitions (see section \ref{sec:level3-408 nm transition saturation}). We form the quantity:
\begin{equation}
L(\theta,f_B;x_i)=\frac{1}{2}\frac{R_0^{[2]'}-R_{-1}^{[2]'}-R_{+1}^{[2]'}}{R_0^{[2]'}+R_{-1}^{[2]'}+R_{+1}^{[2]'}}.
\end{equation}
\noindent This parameter is a function of $\theta$, the magnetic field flipping parameter $f_B$, and all apparatus imperfections (i.e. field imperfections and $h$), which we label as $x_i$. When $x_i\rightarrow0$ then $L(\theta,f_B)=-(1/2)\cos2\theta$, and $L=0$ for $\theta=\pm\pi/4$. We adjust the input to the PBC polarization angles for an $L\approx0$ reading (to within $1\cdot10^{-3}$), and use the measured values of $L$ to determine the actual $\theta_{+}$ and $\theta_{-}$ angles. For a given angle, we average measurements made for both polarities of the magnetic field ($f_B=\pm1$):
\begin{equation}
\label{eq:Lplusminusexact}
\overline{L}_{\pm}=\frac{1}{2}[L(\theta_{\pm},f_B=+1;x_i)+\\L(\theta_{\pm},f_B=-1;x_i)].
\end{equation}
\noindent We use an approximate formula to relate  $\overline{L}_{\pm}$ to $\theta_{\pm}$, that is derived by series expansion of (\ref {eq:Lplusminusexact}) in the small parameters $x_i$, and in $\theta_{\pm}$ around $\pm\pi/4$, respectively:
\begin{equation}
\overline{L}_{\pm}\approx(\pm\theta_{\pm}-\frac{\pi}{4})\mp\frac{b_y^r}{B_z}-\frac{h}{4}.
\end{equation}
\noindent The $\theta_{\pm}$ angles corresponding to measured $\overline{L}_{\pm}$ values are given by:
\begin{equation}
\label{eq:ExpandedThetaPlusMinus}
\theta_{\pm}\approx\pm\frac{\pi}{4}\pm \overline{L}_{\pm}+\frac{b_y^r}{B_z}\pm\frac{h}{4}.
\end{equation}
\noindent We see from (\ref{eq:ExpandedThetaPlusMinus}) that $\theta_{\pm}$ can only be determined with an offset $b_y^r/B_z$ (estimated to be as large as a few parts per 10$^3$), which we do not have an accurate way to measure in the current apparatus [we do make a correction to $\theta_{\pm}$ to account for the contribution of the parameter $h$ present in (\ref{eq:ExpandedThetaPlusMinus})]. This offset, however, does not affect the determination of the parameter $p_{\theta}$ (\ref{eq:pTheta}), used to calibrate the PV data. To show this, we expand $p_{\theta}=\cot\theta_+-\cot\theta_-$ around $\theta_+=\pi/4$ and $\theta_-=-\pi/4$:
\begin{equation}
p_{\theta}\approx 2(1+\frac{\pi}{2}-\theta_++\theta_-),
\end{equation}
\noindent or, with use of (\ref{eq:ExpandedThetaPlusMinus}):
\begin{equation}
p_{\theta}\approx2(1-\overline{L}_+-\overline{L}_--\frac{h}{2}).
\end{equation}
\noindent We see that $p_{\theta}$ is independent of the imperfection $b_y^r/B_z$.  

\end{document}